\documentstyle[12pt,epsfig]{article}
\begin{document}
\thispagestyle{empty}
\begin{titlepage}
\begin{center}
{ \large \bf   SUPERSYMMETRY IN PARTICLE PHYSICS: \\[5mm] THE
RENORMALIZATION GROUP VIEWPOINT}$^\dagger$
 \vspace{1cm}

{\bf D. I.~Kazakov } \\[5mm]

{\it BLTP, JINR, Dubna and ITEP, Moscow
 \\ e-mail: kazakovd@thsun1.jinr.ru}
 \end{center}
\vspace{1cm}

  \leavevmode
  \epsfxsize=4cm
\hspace*{5cm} \epsffile{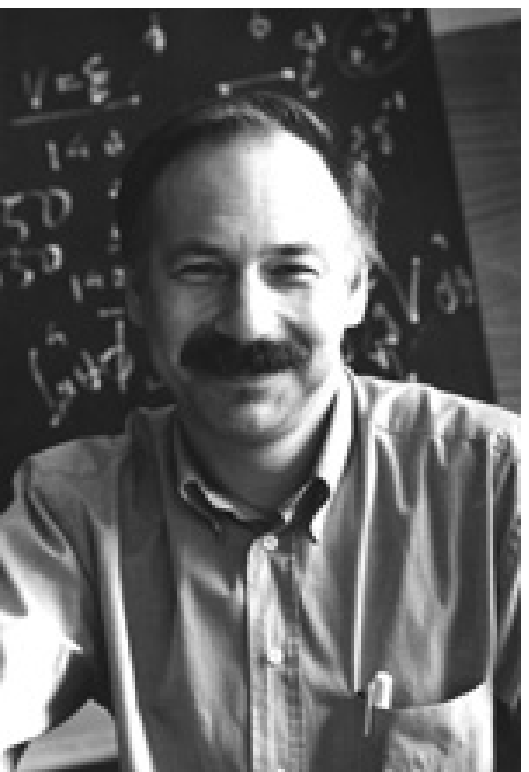}

\vspace{2cm}

\begin{center}
  Abstract
\end{center}

An attempt is made to present modern hopes to find manifestation
of supersymmetry, a new symmetry that relates  bosons and
fermions, in particle physics from the point of view of
renormalization group flow. The Standard Model of particle
interactions is briefly reviewed, the main notions of
supersymmetry are introduced. In more detail the RG flow in the
Minimal Supersymmetric Standard Model is considered. Predictions
of particle spectrum are obtained based on the RG fixed points.
\vspace{1cm}

---------------------------------------------

$^\dagger$  Review talk given at the Conference ``Renormalization
Group at the Turn of the Millennium'', Taxco, Mexico, January 1999

\pagebreak
 \tableofcontents \vglue 0.34cm {\bf References} \hfill
{\bf 41}

\end{titlepage}
\renewcommand{\thesubsection}{\thesection.\arabic{subsection}}
\renewcommand{\thesubsubsection}{\thesubsection.\arabic{subsubsection}}
\renewcommand{\theequation}{\thesection.\arabic{equation}}
\section{Introduction}
\setcounter{equation} 0

\subsection{The Standard Model}

The Standard Model (SM) of fundamental interactions describes {\em
strong, weak} and {\em electromagnetic } interactions of
elementary particles  \cite{SM}. It is based on a {\em gauge
principle}, according to which all the forces of Nature are
mediated by an exchange of the gauge fields of a corresponding
local symmetry group. The symmetry group of the SM is
\begin{equation}
SU_{colour}(3)\otimes SU_{left}(2)\otimes U_{hypercharge}(1),
\label{a}
\end{equation}
whereas the field content is the following:

\vspace{0.4cm} \noindent\underline{{\em Gauge sector} :\ \ Spin =
1} \par \vspace{0.4cm}

The gauge bosons are spin 1 vector particles belonging to the
adjoint representation of the group (\ref{a}). Their quantum
numbers with respect to $SU(3)\otimes SU(2)\otimes U(1)$ are: $$
\begin{array}{ccccc}
\mbox{\em gluons}& G_\mu^a :& \ (\underline{8},\underline{1},0)& \
\ SU_c(3)& \ \ g_{s},\\ {\begin{array}{cc} \mbox{\em
intermediate}\\ \mbox{\em weak bosons} \end{array}}& W_\mu^i :& \
(\underline{1}, \underline{3},0)& \ \ SU_L(2)& \ \ g,\\ \mbox{\em
abelian boson}& B_\mu :& \ (\underline{1},\underline{1},0)& \ \
U_Y(1)&\ \ g',
\end{array} $$
where the coupling constants are usually denoted by $g_s$, $g$ and
$g'$, respectively.

\vspace{0.4cm} \noindent\underline{{\em Fermion sector} : \ \ Spin
= 1/2} \par \vspace{0.4cm}

The matter fields are fermions belonging to the fundamental
representation of the gauge group. These are believed to be quarks
and leptons of at least of three generations. The SM is left-right
asymmetric. Left-handed and
 right-handed fermions have different quantum numbers:
$$
\begin{array}{cccccccc}
\mbox{\em quarks} \ & & & & & & & \\
 Q_{\alpha L}^i = & \left(\begin{array}{c}U_\alpha^i\\D_\alpha^i
\end{array} \right)_L & = & \left(\begin{array}{c}u^i\\d^i \end{array}\right)_L ,
& \left(\begin{array}{c}c^i\\s^i \end{array}\right)_L,&
\left(\begin{array}{c}t^i
\\b^i \end{array} \right)_L,& \ldots & (\underline{3},\underline{2},1/3)\\
  &  U_{\alpha R}^i & = & u_{iR},& c_{iR},& t_{iR},& \ldots &
(\underline{3}^*, \underline{1}, 4/3)\\ &  D_{\alpha R}^i & = &
d_{iR},&
  s_{iR},& b_{iR},& \ldots &(\underline{3}^*, \underline{1}, -2/3)\\
\mbox{\em leptons} \ &  L_{\alpha L} & = &
\left(\begin{array}{c}\nu_e\\ e \end{array} \right)_L, &
\left(\begin{array}{c}\nu_\mu\\ \mu  \end{array} \right)_L, &
\left(\begin {array}{c}\nu_\tau\\ \tau \end{array} \right)_L,&
\ldots & (\underline{1}, \underline{2},-1)\\ &  E_{\alpha R} & = &
e_R,&
  \mu_R,& \tau_R,& \ldots & (\underline{1},\underline{1}, -2) \end{array}  $$
$i = 1,2,3$ - colour, $\alpha = 1,2,3,\ldots$ - generation.

\vspace{0.4cm} \noindent\underline{{\em Higgs sector} : \ \ Spin =
0} \par \vspace{0.4cm}

 In the minimal version of the SM there is  one doublet of Higgs
 scalar fields
\begin{equation}
H  =  \left(\begin{array}{c}H^{0}\\ H^- \end{array} \right) \ \ \
\ (\underline{1}, \underline{2},-1),
\end{equation}
which is introduced in  order to give masses to quarks, leptons
and intermediate weak bosons via spontaneous  breaking of
electroweak symmetry.

In Quantum Field Theory framework the SM is described by the
following Lagrangian
\begin{equation}
{\cal L} ={\cal L}_{gauge} + {\cal L}_{Yukawa} + {\cal L}_{Higgs},
\label{SM}
 \end{equation}
\begin{eqnarray}
{\cal L}_{gauge} & = & -\frac{1}{4} G_{\mu\nu}^aG_{\mu\nu}^a -
\frac{1}{4} W_{\mu\nu}^iW_{\mu\nu}^i -\frac{1}{4}
B_{\mu\nu}B_{\mu\nu}\\ & &  +
 i\overline{L}_{\alpha}\gamma^{\mu}D_{\mu}L_{\alpha} +
i\overline{Q}_{\alpha}\gamma^{\mu}D_{\mu}Q_{\alpha} +
i\overline{E}_{\alpha} \gamma^{\mu}D_{\mu}E_{\alpha} \nonumber \\
 & &  + i\overline{U}_{\alpha}\gamma^{\mu}D_{\mu}U_{
\alpha} + i\overline{D}_{\alpha}\gamma^{\mu}D_{\mu}D_{\alpha} +
(D_{\mu}H)^{\dagger}(D_{\mu}H),  \nonumber
\end{eqnarray}
where
\begin{eqnarray*}
G_{\mu\nu}^a & = &
\partial_{\mu}G_{\nu}^a-\partial_{\nu}G_{\mu}^a+g_{s}
f^{abc}G_{\mu}^bG_{\nu}^c,\\ W_{\mu\nu}^i & = &
\partial_{\mu}W_{\nu}^i-\partial_{\nu}W_{\mu}^i+g\epsilon
^{ijk}W_{\mu}^jW_{\nu}^k,\\ B_{\mu\nu} & = &
\partial_{\mu}B_{\nu}-\partial_{\nu}B_{\mu},\\ D_{\mu}L_{\alpha} &
= &
(\partial_{\mu}-i\frac{g}{2}\tau^iW_{\mu}^i+i\frac{g'}{2}B_{\mu})L_{\alpha},\\
D_{\mu}E_{\alpha} & = & (\partial_{\mu}+ig'B_{\mu})E_{\alpha},\\
D_{\mu}Q_{\alpha} & = &
(\partial_{\mu}-i\frac{g}{2}\tau^iW_{\mu}^i-i\frac{g'}{6}B_{\mu}-
i\frac{g_s}{2}\lambda^aG_{\mu}^a)Q_{\alpha},\\ D_{\mu}U_{\alpha} &
= &
(\partial_{\mu}-i\frac{2}{3}g'B_{\mu}-i\frac{g_s}{2}\lambda^aG_{\mu}^a)
U_{\alpha},\\ D_{\mu}D_{\alpha} & = &
(\partial_{\mu}+i\frac{1}{3}g'
B_{\mu}-i\frac{g_s}{2}\lambda^aG_{\mu}^a)D_{\alpha}.
\end{eqnarray*}
\begin{equation}
{\cal L}_{Yukawa} = y_{\alpha\beta}^L\overline{L}_{\alpha}E_{\beta}H +
y_{\alpha \beta}^D\overline{Q}_{\alpha}D_{\beta}H +
y_{\alpha\beta}^U\overline{Q}_{\alpha} U_{\beta}\tilde{H} + h.c.,
\label{yuk}
\end{equation}
where $\tilde{H}=i\tau_2H^{\dagger}$.
\begin{equation}
{\cal L}_{Higgs} = - V = m^2H^{\dagger}H -
\frac{\lambda}{2}(H^{\dagger}H)^2.\label{hig}
\end{equation}
Here $\{y\}$ are the Yukawa and $\lambda$ is the Higgs coupling
constants, both di\-men\-si\-on\-less, and $m$ is the only
dimensional mass parameter\footnote{We use the usual for particle
physics units $c=\hbar=1$}.

The Lagrangian of the SM contains the following set of free
parameters:
\begin{itemize}
     \item 3 gauge couplings $g_s, g, g' $;
  \item 3 Yukawa matrices $y_{\alpha\beta}^L, y_{\alpha\beta}^D, y_{\alpha\beta}^U
$;
     \item Higgs coupling  constant $\lambda$;
     \item Higgs mass parameter $m^2$;
     \item number of matter fields (generations).
\end{itemize}

All the particles obtain their masses due to spontaneous breaking of
$SU_{left}(2)$ symmetry group via a non-zero vacuum expectation value
(v.e.v.) of the Higgs field
\begin{equation}
<H> = \left(\begin{array}{c}v\\ 0\end{array}\right),\ \ \ \
v=m/\sqrt{\lambda}.
\end{equation}
As a result  the gauge group of the SM is spontaneously  broken down to
$$SU_c(3)\otimes SU_L(2)\otimes U_Y(1) \Rightarrow SU_c(3)\otimes
U_{EM}(1).$$ The physical weak intermediate bosons are the linear
combinations of the gauge ones
\begin{equation} W_{\mu}^{\pm} =  \frac{W_{\mu}^1\mp iW_{\mu}^2}{\sqrt{2}},\
\ \ \ Z_{\mu} = -\sin{\theta_W}B_{\mu} + \cos{\theta_W}W_{\mu}^3
\end{equation}
with masses
\begin{equation}
m_W=\frac{1}{\sqrt{2}}gv,\ \ \ \  m_Z=m_W/\cos{\theta_W},\ \ \ \
\tan{\theta_W}=g'/g, \label{Z}
\end{equation}
while the photon field
\begin{equation}
\gamma_{\mu} = \cos{\theta_W}B_{\mu} + \sin{\theta_W}W_{\mu}^3
\end{equation}
remains massless.

The matter fields acquire masses proportional to the corresponding
Yukawa couplings:
\begin{equation}
M_{\alpha\beta}^u = y_{\alpha\beta}^uv,\ M_{\alpha\beta}^d =
y_{\alpha\beta}^dv,\ M_{\alpha\beta}^l = y_{\alpha\beta}^lv, \ m_H
= \sqrt{2}m. \label{mass}
\end{equation}
Explicit mass terms in the Lagrangian are forbidden because they
are not $SU_{left}(2)$ symmetrical and would destroy the
renormalizability of the Standard Model.

The SM has been constructed as a result of numerous efforts both theoretical
and experimental.  At present the SM is extraordinary successful, the
achieved accuracy of its predictions corresponds to the experimental data
within 5 \% \cite{test-sm}. All the particles except for the Higgs boson
have been discovered experimentally.

However the SM has its natural drawbacks and unsolved problems.
Among them are:
\begin{itemize}
\item large number of free parameters,
\item formal unification of strong and electroweak interactions,
\item the Higgs  boson has not yet been observed and it is not clear
whether it is fundamental or composite,
\item the problem of CP-violation is not well understood including
CP-violation in strong interaction,
\item flavour mixing and the number of generations are arbitrary,
\item the origin of the mass spectrum is unclear.
\end{itemize}
The answer to these problems lies beyond the SM.

\subsection{RG flow in the Standard Model}

The question is: how to go beyond the SM? Apparently we are
talking about new particles, new structures, new interactions  and
new symmetries. The answer is not obvious. We describe below one
of the options and assume that at high energies (small distances)
fundamental interactions possess wider symmetries, in particular,
new kind of symmetry, symmetry between bosons and fermions, called
sypersymmetry. This is a wide subject by itself, however, we would
like to look at it from the point of view of renormalization
group. Let us try to go along the road offered by the
renormalization group flow.

 Let us take the Lagrangian of the SM and see
what happens with its parameters (the couplings) when the energy
scale increases. As it follows from eq.(\ref{a}) one has three
gauge couplings corresponding to $SU(3)_c \times SU(2)_L \times
U(1)_Y$ gauge groups, $g_s, g$ and $g'$, respectively. In what
follows it will be more useful to consider the set
$\{g_1,g_2,g_3\}=\{\sqrt{3/5}g',g,g_s\}$. Besides, there are three
Yukawa couplings $y_U, y_D$ and $y_L$, which are $3\times3$
matrices in the generation space and one Higgs coupling $\lambda$.

To simplify the picture we consider only the third generation
Yukawa couplings replacing the Yukawa matrices by their $(3,3)$
elements. Then the RG equations in the leading one-loop order are
\cite{RGSM} :
\begin{eqnarray}
\frac{d{\mbox a}_i}{dt}& =&  b_i{\mbox a}_i^2, \ \ \ \ \
t=\log(Q^2/\mu^2),\label{alpha} \\
 \frac{dY_t}{dt} & = &-  Y_t\left( 8{\mbox a}_3 +
\frac{9}{4}{\mbox a}_2 + \frac{17}{20}{\mbox a}_1-\frac{9}{2}Y_t\right),
\nonumber\\
 \frac{dY_b}{dt} & =&-  Y_b\left(
8{\mbox a}_3 + \frac{9}{4}{\mbox a}_2 +
\frac{1}{4}{\mbox a}_1-\frac{3}{2}Y_t\right),\label{yukawa} \\
 \frac{dY_\tau}{dt} & =& -  Y_\tau\left(
  \frac{9}{4}{\mbox a}_2 + \frac{9}{4}{\mbox a}_1-3Y_t\right),\nonumber \\
 \frac{d\lambda }{dt} &=&
   \frac{9}{8}{\mbox a}_2^2 + \frac{9}{20} {\mbox a}_2{\mbox a}_1 +
  \frac{27}{200}{\mbox a}_1^2 - \frac{9}{2}{\mbox a}_2\lambda
   -\frac{9}{10}{\mbox a}_1\lambda  + 6 \lambda Y_t - 6 Y_t^2 + 6 \lambda^2,
      \label{lambda}
\end{eqnarray}
where to simplify the formulas we have used the notation
$${\mbox a}_i\equiv \frac{\alpha_i}{4\pi} \equiv
\frac{g_i^2}{16\pi^2}, \ \ \ (i=1,2,3), \ \ \ Y_k\equiv
 \frac{(y^k_{33})^2}{16\pi^2}, \ \ \ (k=t,b,\tau).$$
For the SM the coefficients $b_i$ are:
\begin{equation}
b_i=\left( \begin{array}{r} b_1 \\ b_2 \\b_3 \end{array} \right)
   =
\left( \begin{array}{r}           0    \\
                             - 22 / 3  \\
                                -11    \end{array} \right) +N_{Fam}
\left( \begin{array}{r}         4 / 3  \\
                                4 / 3  \\
                                4 / 3  \end{array} \right) + N_{Higgs}
\left( \begin{array}{r}         1 / 10 \\
                                1 / 6  \\
                                  0    \end{array} \right) .
\label{regeq1}
\end{equation}
Here  $N_{Fam}$ is the number of generations of matter multiplets
and $N_{Higgs}$ is the number of Higgs doublets. We use
$N_{Fam}=3$ and $N_{Higgs}=1$ for the minimal SM.

 The initial conditions for the couplings can be taken at some scale which
is experimentally favorable. Thus, for the gauge couplings one has
precise measurement at the Z-boson mass scale obtained at LEP
accelerator \cite{LEP}
\begin{equation}
\alpha_1(M_Z)=0.017,\ \ \alpha_2(M_Z)=0.034, \ \
\alpha_3(M_Z)=0.118\pm 0.005.
\end{equation}
As for the Yukawa couplings, they are related to the running quark
masses by eq.(\ref{mass}), where $v$ is the vacuum expectation
value of the Higgs field. It can be calculated, for instance,
from the Z-boson mass according to eq.(\ref{Z}) and is equal to
$v=174.1$ GeV. Thus, knowing the quark masses one can find the
values of the Yukawa couplings. One should however distinguish
between the running and the pole quark masses which are determined
experimentally.

\begin{figure}[htb]
 \begin{flushleft}
     \leavevmode
    \epsfxsize=8cm
   \epsffile{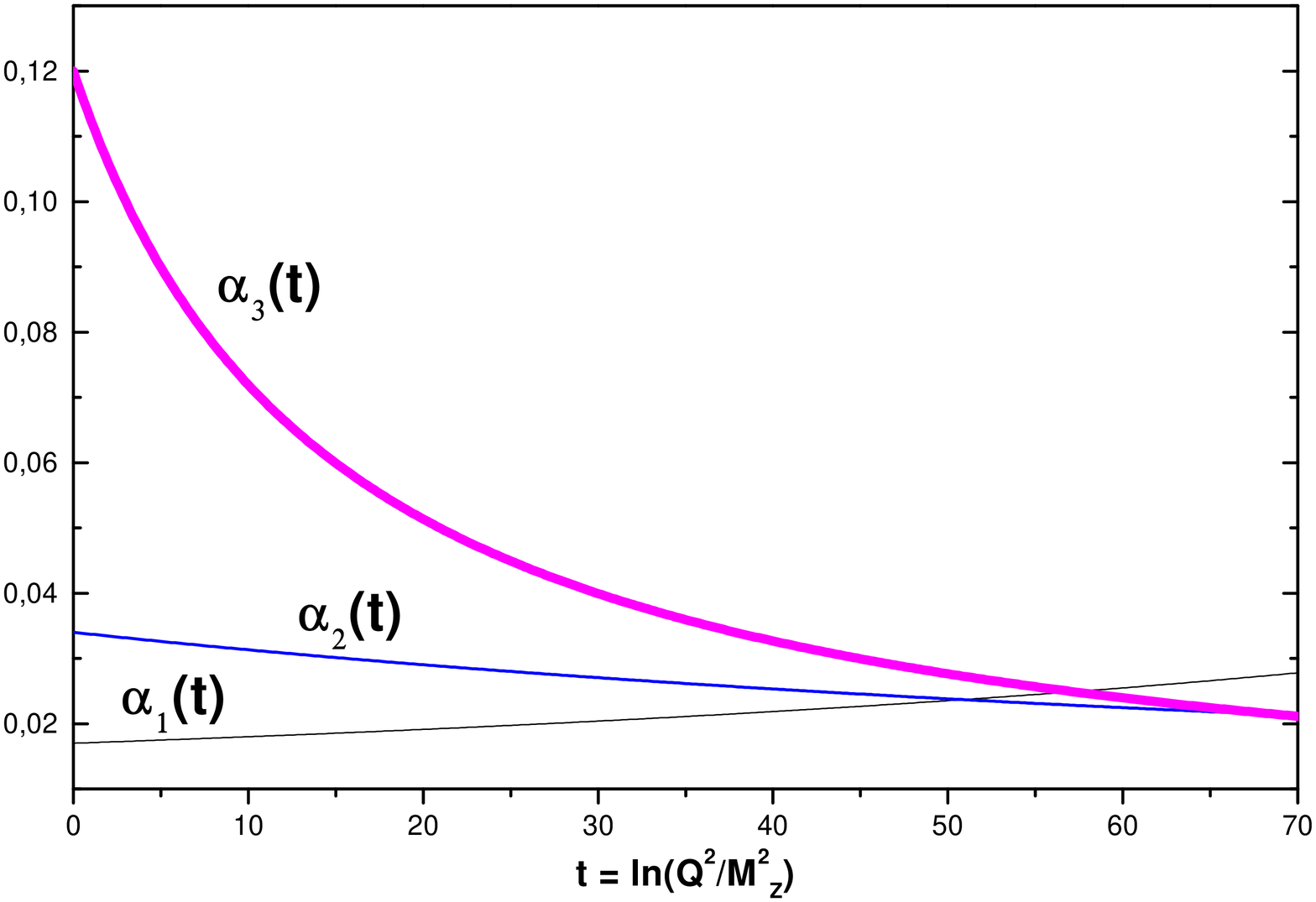} 
  \end{flushleft}
\vspace{-6cm} \hspace{7.8cm}
  \leavevmode
  \epsfxsize=8cm
\epsffile{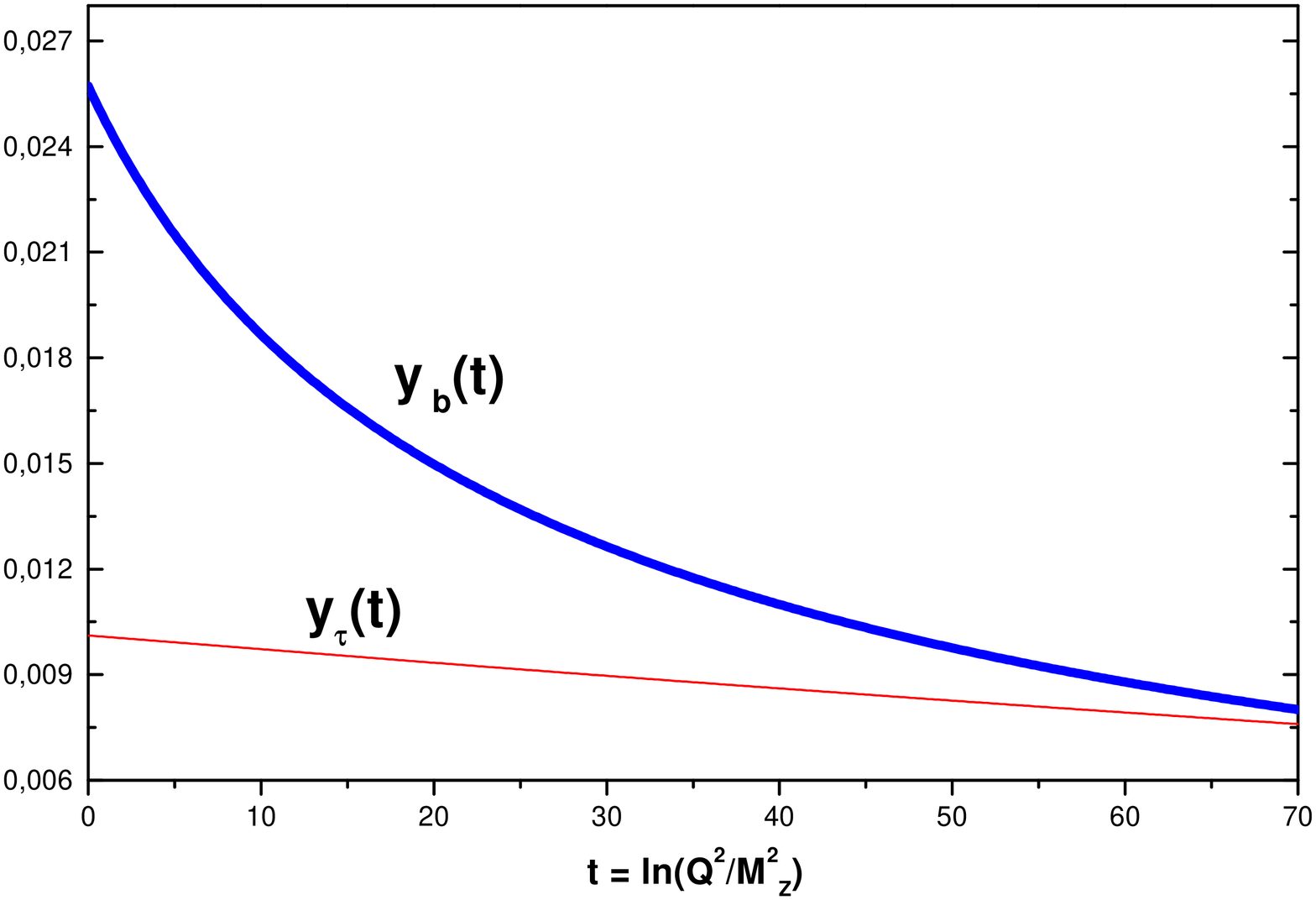}
\caption{Evolution of the gauge (left)  and Yukawa couplings
(right) in the Standard Model. The width of the curves corresponds
to the experimental error.} \label{smcoup}
\end{figure}

Having all this in mind and solving eqs.(\ref{alpha}-\ref{yukawa})
one has the following qualitative picture (see Fig.\ref{smcoup}).
The behaviour of the Higgs quartic coupling $\lambda$ strongly
depends on initial conditions which are unknown in this case. We
return to this subject later.

 The qualitative picture presented in
Fig.\ref{smcoup} contains an obvious unification pattern. The
three gauge couplings are seem to unify at energy of the order of
$10^{15}-10^{16}$ GeV and so do the  Yukawa couplings  $Y_b$ and
$Y_\tau$. What does it mean? The usual answer is given in the
framework of the Grand Unification hypothesis \cite{GUT}: three
gauge interactions are the three branches of a single gauge
interaction described by a simple gauge group with a single
coupling. All quarks and leptons belong to some representation of
this group. This explains the equality of gauge and (some) Yukawa
couplings at the unification scale.

The GUT hypothesis has many far reaching consequences, however, one can see
that the unification scale is very high. This is not only difficult to check
experimentally, but creates a big problem, called the {\em hierarchy}
problem.

The point is that in a theory with two so very different scales:
$M_W \sim 10^2$ GeV and $M_{GUT}\sim 10^{16}$ GeV, it is very
difficult both  to achieve this hierarchy of $10^{14}$ in a
natural way and to preserve it against the radiative corrections.

Indeed, due to  modern point of view, the mass scales in the SM
and in GUT are defined by vacuum expectation values of the scalar
fields, called the Higgs fields. Non-zero v.e.v.'s of these fields
lead to spontaneous breaking of the corresponding gauge symmetry
and provide masses to all the particles. So, we have at least two
scalar particles with the masses of the order of $10^2$ and
$10^{16}$ GeV.  However, these masses obtain the radiative
corrections proportional to the masses of the interacting
particles. Due to inevitable interaction between the light and
heavy  fields the radiative corrections to the light Higgs mass
are proportional to the heavy one $$\delta m^2 \sim g^2 M^2,$$
where $g$ is some coupling. Assuming $m \sim 10^2\ GeV, \ M \sim
10^{16} \ GeV, \ g \sim 0.1$, one gets the radiative correction
which is $10^{13}$ times bigger than the mass itself.
 This correction obviously spoils the hierarchy unless  it is
canceled. A  cancellation with a precision $\sim 10^{-13}$ needs a
very accurate fine tuning of the coupling constants.

 Solution
to the fine-tuning problem has been found in the framework of a
revolutionary hypothesis: the existence of a new type of symmetry,
the symmetry between bosons and fermions, called supersymmetry.

\section{Supersymmetry}
\setcounter{equation} 0

\subsection{Motivations of SUSY}

Supersymmetry or fermion-boson symmetry has not yet been observed
in Nature. This is a purely theoretical invention~\cite{super}.
Its validity in particle physics follows from common belief in
unification. The {\em general idea} is a unification of all forces
of Nature. It defines the {\em strategy} : increasing unification
towards smaller distances up to $l_{Pl} \sim 10^{-33} $ cm
including quantum gravity. However, the graviton has spin 2, while
the other gauge bosons (photon, gluons, $W$ and $Z$ weak bosons)
have spin 1. Unification of spin 2 and spin 1 gauge forces within
unique algebra is forbidden due to the no-go theorems for any
symmetry but SUSY.

If $Q$ is a generator of SUSY algebra, then $$ Q| boson> = |
fermion> \ \ \ \mbox{and} \ \ \ Q| fermion> = | boson> .$$ Hence
starting with the graviton spin 2 state and acting by SUSY
generators we get the following chain of states $$spin \ 2\ \
\rightarrow \ \ spin \ 3/2 \ \ \rightarrow \ \ spin \ 1 \ \
\rightarrow \ \ spin \ 1/2 \ \ \rightarrow \ \ spin \ 0 .$$ Thus,
a partial unification of matter (fermions) with forces (bosons)
naturally arises out of an attempt to unify gravity with the other
interactions.

The {\em uniqueness} of SUSY is due to a strict mathematical
statement that algebra of SUSY is the {\em only} graded (i.e.
containing anticommutators as well as commutators) Lie algebra
possible within relativistic field theory \cite{theorem}.

The other motivation of SUSY is the solution of the hierarchy
problem mentioned above. At the moment supersymmetry is the
 only known way to achieve the cancellation of quadratic
terms in radiative corrections (also known as the cancellation of
the quadratic divergences). Moreover, SUSY automatically cancels
quadratic corrections in all orders of perturbation theory
\cite{West}.

\subsection{Global SUSY: algebra and representations}

As can be easily seen, supersymmetry trans\-for\-ma\-tions differ
from ordinary global transformations as far as they convert bosons
into fermions and vice versa. Indeed if we symbolically write SUSY
transformation as $$ \delta B = \varepsilon \cdot f,$$ where $B$
and $f$ are boson and fermion fields, respectively, and
$\varepsilon $ is an infinitesimal trans\-for\-mation parameter,
then from the usual (anti)commutation relations for (fermions)
bosons $$ \{f,f\} = 0, \ \ \ \  [B,B] = 0  $$ we immediately find
$$\{ \varepsilon , \varepsilon \} = 0. $$ This means that all the
generators of SUSY must be {\em fermionic}, i.e. they must change
the spin by a half-odd amount and change the statistics.

Combined with the usual Poincar\'e and internal symmetry algebra
the Super-Poincar\'e Lie algebra contains additional generators
\cite{super}
\begin{equation}  \begin{array}{l}
 {[} Q_{\alpha}^i , P_{\mu} {]} =  {[} \bar Q_{\dot \alpha}^i , P_{\mu} {]} = 0, \\
 {[} Q_{\alpha}^i , M_{\mu \nu} {]}  =
\frac{1}{2} (\sigma_{\mu \nu})_{\alpha}^{\beta}Q_{\beta}^i , \ \ \
\ {[}\bar Q_{\dot \alpha}^i ,M_{\mu \nu}{]} = - \frac{1}{2} \bar
Q_{\dot \beta}^i (\bar \sigma_{\mu \nu})_{\dot \alpha} ^{\dot
\beta} , \\
 {[} Q_{\alpha}^i , B_r {]}  =  (b_r)_{j}^i Q_{\alpha}^j , \ \ \ {[}
 \bar Q_{\dot \alpha}^i , B_r {]} = - \bar Q_{\dot \alpha}^j (b_r)_j^i , \\
 \{ Q_{\alpha}^i , \bar Q_{\dot \beta}^j \}
  =   2 \delta^{ij} (\sigma ^{\mu})_{\alpha \dot \beta }P_{\mu} , \\
 \{ Q_{\alpha}^i , Q_{\beta}^j \}  =   2 \epsilon_{\alpha \beta}Z^{ij} ,
\ \ \ Z_{ij} = a_{ij}^r b_r , \ \ \ \ Z^{ij} = Z_{ij}^+ , \\
 \{ \bar Q_{\dot \alpha}^i , \bar Q_{\dot \beta}^j \}  =  - 2 \epsilon
_{\dot \alpha \dot \beta}Z^{ij} , \ \ \ {[}Z_{ij} , anything {]} =
0 , \\
 \alpha , \dot \alpha, \beta , \dot \beta   =  1,2 \ \ \ \
i,j = 1,2, \ldots , N .  \end{array}  \label{f}
\end{equation}
Here $P_{\mu}$ and $M_{\mu \nu}$  are four-momentum and angular momentum
operators respectively, $B_r$ are internal symmetry generators, $Q^i$ and
$\bar Q^i$ are spinorial SUSY generators and $Z_{ij}$ are the so-called
central charges. $\alpha , \dot \alpha, \beta , \dot \beta $ are  spinorial
indices. In the simplest case one has one spinor generator $Q_\alpha$ (and the
conjugated one $\bar Q_{\dot{\alpha}}$) that corresponds to an ordinary or
N=1 sypersymmetry. When $N>1$ one has an extended sypersymmetry.

An elegant formulation of supersymmetry transformations and
invariants can be achieved in the framework of superspace
\cite{sspace}.
 Superspace differs from the ordinary Euclidean (Minkowski)
space by addition of two new coordinates, $\theta_{\alpha}$ and
$\bar \theta_{\dot \alpha}$, which are grassmannian,
 i.e. anti\-com\-muting, variables
$$\{ \theta_{\alpha}, \theta_{\beta} \} = 0 , \ \ \{\bar \theta_{\dot \alpha},
\bar \theta_{\dot \beta} \} = 0, \ \ \theta_{\alpha}^2 = 0,\ \
\bar \theta_{\dot \alpha}^2=0, \ \ \alpha,\beta, \dot\alpha, \dot\beta =1,2.$$
Thus, we go from space to superspace
$$\begin{array}{cc} Space & \ \Rightarrow \ \ Superspace \\
x_{\mu} & \ \ \ \ \ \ \ x_{\mu}, \theta_{\alpha} , \bar \theta_
{\dot \alpha}
\end{array}$$
 A SUSY group element can be constructed in
superspace in the same way as an ordinary translation in the usual
space $$G(x,\theta ,\bar \theta ) = e^{\displaystyle
i(-x^{\mu}P_{\mu} + \theta Q + \bar \theta \bar Q)}.$$ It leads to
the supertranslation in superspace
\begin{equation}
\begin{array}{ccl}
x_{\mu} & \rightarrow & x_{\mu} + i\theta \sigma_{\mu} \bar \varepsilon
 - i\varepsilon \sigma_{\mu} \bar \theta, \\
\theta & \rightarrow & \theta + \varepsilon , \\ \bar \theta & \rightarrow &
\bar \theta + \bar \varepsilon , \end{array}
\end{equation}
where $\varepsilon $ and $\bar \varepsilon $ are grassmannian
transformation parameters. Taking them to be local or space-time
dependent one gets   local translation. And the theory that is
invariant under local translations is general relativity. Thus
local supersymmetry is just the theory of gravity or supergravity
\cite{SUGRA}. This way following the gauge principle one gets a
unified theory of all four interactions known as SUGRA theory.

To define the fields  on a superspace  consider  representations
of the Super-Poincar\'e group (\ref{f}) \cite{WessB}. The simplest
one is a scalar superfield $F(x,\theta , \bar \theta )$ which is
SUSY invariant. Its Taylor expansion in $\theta$ and $\bar \theta
$ has only several terms due to the nilpotent character of
grassmannian parameters. However, this superfield  is a reducible
representation of SUSY. To get an irreducible one, we define a
{\em chiral} superfield which obeys the equation
\begin{equation}
\bar D F = 0 , \ \ \ \ \mbox{where} \ \bar D = -\frac{\partial}{
\partial \overline{ \theta}} - i \theta \sigma^{\mu}
\partial_{\mu}. \label{k} \end{equation}
Its Taylor expansion
looks like ($y =x + i\theta \sigma \bar \theta  $)
\begin{eqnarray}
\Phi (y, \theta ) & = & A(y) + \sqrt{2} \theta \psi (y) + \theta
 \theta F(y) \nonumber \\
  & = & A(x) + i\theta \sigma^{\mu} \bar \theta \partial_{\mu}A(x)
 + \frac{1}{4} \theta \theta \bar \theta \bar \theta \Box A(x) \nonumber\\
  & + & \sqrt{2} \theta \psi (x) - \frac{i}{\sqrt{2}} \theta
\theta \partial_{\mu} \psi (x) \sigma^{\mu} \bar \theta + \theta
 \theta F(x)\label{field}
\end{eqnarray}
The coefficients are ordinary functions of $x$ being the usual
fields. They are called the {\em components} of a superfield. In
eq.(\ref{field}) one has 2 bosonic (complex scalar field $A$) and
2 fermionic (Weyl spinor field $\psi$) degrees of freedom. The
componet fields $A$ and $\psi$ are called the {\em superpartners}.
The field $F$ is  an {\em auxiliary} field, it has the ``wrong''
dimension and has no physical meaning. It is needed to close the
algebra (\ref{f}). One can get rid of the auxiliary fields with
the help of equations of motion.

Thus a superfield contains an equal number of bosonic and
fermionic degrees of freedom. Under SUSY transformation they
convert one into another.

The product of chiral superfields $\Phi^2 , \Phi^3 $, etc is also
a chiral superfield, while the product of chiral and antichiral
ones $\Phi^+ \Phi$ is a general superfield.

To construct the gauge invariant interactions, we will need a real
vector superfield $V = V^+$. It is not chiral but rather a general
superfield. Its expansion over $\theta$ and $\bar\theta$ looks
like
\begin{eqnarray}
V(x, \theta, \bar \theta) & = & C(x) + i\theta \chi (x) -i\bar
\theta \bar \chi (x)  \nonumber \\
 & + & \frac{i}{2} \theta \theta [M(x) + iN(x)] - \frac{i}{2} \bar
 \theta \bar \theta [M(x) - iN(x)]  \nonumber \\
 & - & \theta \sigma^{\mu} \bar \theta v_{\mu}(x) + i \theta \theta
\bar \theta [\bar \lambda (x) + \frac{i}{2}\bar \sigma^{\mu}
\partial _{\mu} \chi (x)]  \nonumber \\
 & - & i\bar \theta \bar \theta \theta [\lambda + \frac{i}{2}
\sigma^{\mu} \partial_{\mu} \bar \chi (x)] + \frac{1}{2} \theta
\theta \bar \theta \bar \theta [D(x) + \frac{1}{2}\Box C(x)].
 \label{p} \end{eqnarray}
The physical degrees of freedom corresponding to a real vector
superfield are the vector gauge field $v_{\mu}$ and its
superpartner the Majorana four component spinor field made of two
Weyl spinors $\lambda$ and $\bar \lambda$. All other components
are unphysical and can be eliminated. Thus we again have an equal
number of bosonic and fermionic degrees of freedom.

One can choose a gauge (Wess-Zumino gauge) where $C = \chi = M = N
=0 $, leaving us with the physical degrees of freedom except for
the auxiliary field $D$. In this gauge
\begin{eqnarray}
V & = & - \theta \sigma^{\mu} \bar \theta v_{\mu}(x) + i \theta
\theta \bar \theta \bar \lambda (x) -i\bar \theta \bar \theta
\theta \lambda (x) + \frac{1}{2} \theta \theta \bar \theta \bar
\theta D(x) , \nonumber\\ V^2 & = & - \frac{1}{2} \theta \theta
\bar \theta \bar \theta v_{\mu}(x)v^{\mu}(x) , \nonumber\\ V^3 & =
& 0, \ \ \ etc. \end{eqnarray} One can define also a field
strength tensor (as analog of $F_{\mu \nu}$ in gauge theories)
\begin{equation}
W_{\alpha}  = - \frac{1}{4} \bar D^2 e^{-V}D_{\alpha} e^V , \ \ \
\  \bar W_{\dot \alpha}  =  - \frac{1}{4} D^2 e^{-V} \bar
D_{\alpha} e^V, \label{w}
\end{equation}
which is needed to construct  gauge invariant Lagrangians.

\subsection{SUSY Lagrangians}

In the superfield notation SUSY invariant Lagrangians are the
 polynomials of superfields. Having in mind that for component
fields we should have the ordinary terms, the general SUSY
invariant Lagrangian has the form \cite{WessB, West, sspace}
\begin{eqnarray}
{\cal L} & = &\int d^2\theta d^2\bar \theta \Phi_i^+ \Phi_i+ \int d^2\theta
 [\lambda_i \Phi_i + \frac{1}{2}m_{ij}\Phi_i \Phi_j + \frac{1}{3} y_{ijk}
\Phi_i \Phi_j \Phi_k] + h.c. \label{l}
\end{eqnarray}
where the first part is a kinetic term and the second one is a
superpotential  ${\cal W}$. Here instead of taking the proper
components we use  an integration over the superspace according to
the rules of grassmannian integration \cite{ber}
 $$\int \ d\theta_\alpha =
0 , \ \ \ \ \int \theta _\alpha\ d\theta _\beta= \delta_{\alpha\beta}. $$
Performing this integration we get in components
\begin{eqnarray}
{\cal L} & = & i\partial_{\mu}\bar \psi_i \bar \sigma^{\mu}\psi_i
+ A_i^{\ast} \Box A_i + F_i^{\ast}F_i \label{20} \\
 & + & [\lambda_i F_i + m_{ij}(A_iF_j - \frac{1}{2}\psi_i \psi_j )
+ y_{ijk}(A_iA_jF_k - \psi_i \psi_j A_k ) + h.c. ] . \nonumber
\end{eqnarray}
or solving the constraints
\begin{eqnarray}
{\cal L} & = & i\partial_{\mu}\bar \psi_i \bar \sigma^{\mu}\psi_i
 + A_i^* \Box A_i - \frac{1}{2}m_{ij}\psi_i \psi_j -
\frac{1}{2}m_{ij}^* \bar \psi_i \bar \psi_j \nonumber \\ & & -
y_{ijk}\psi_i \psi_j A_k - y_{ijk}^* \bar \psi_i \bar \psi_j A_k^*
- V(A_i,A_j),
 \end{eqnarray}
where $V = F_k^* F_k $. Note that because of the renormalizability
constraint $V \leq A^4$ the superpotential should be limited by
${\cal W} \leq \Phi^3 $ as in eq.(\ref{l}).

The gauge field  part of a Lagrangian  is
\begin{eqnarray}
{\cal L} & = & \frac{1}{4}\int d^2\theta W^{\alpha}W_{\alpha}  +
\frac{1}{4}\int d^2\bar \theta \bar W^{\dot \alpha}\bar W_{\dot
\alpha} \nonumber \\
 & = & \frac{1}{2}D^2 - \frac{1}{4}F_{\mu \nu}F^{\mu \nu} -
 i \lambda \sigma^{\mu}D_{\mu}\bar \lambda. \label{29}
\end{eqnarray}
To obtain a gauge-invariant interaction with matter chiral
superfields, consider their gauge transformation  $$\Phi \ \
\rightarrow \ \ e^{-ig\Lambda} \Phi , \ \ \ \Phi^+ \ \ \rightarrow
\ \ \Phi^+ e^{ig\Lambda^+} , \ \ \ V \ \ \rightarrow \ \ V +
i(\Lambda - \Lambda^+) , $$ where $\Lambda $ is a gauge parameter
(chiral superfield).

It is clear now how to construct both  SUSY and gauge invariant interaction
which is equivalent to transition from the usual to covariant derivatives
 $$
\int d^2\theta d^2\bar \theta \Phi_i^+ \Phi_i \Rightarrow \int d^2\theta
d^2\bar \theta \Phi_i^+ e^{gV} \Phi_i . $$

Thus, the form of the Lagrangian  is practically fixed by symmetry
requirements. The only freedom is the field content, the value of the gauge
coupling $g$ ,  the Yukawa couplings $y_{ijk}$ and the masses. This allows one to
construct a SUSY generalization of the SM.

\section{Minimal Supersymmetric Standard Model}
\setcounter{equation} 0

As has been already mentioned, in SUSY theories the number of
bosonic degrees of freedom equals that of fermionic. In the SM one
has 28 bosonic and 90 fermionic degrees of freedom. So the SM is
in great deal non-supersymmetric. Trying to add some new particles
to supersymmetrize the SM, one should take into account the
following observations:
\begin{enumerate}
\item There are no fermions with quantum numbers of the
gauge bosons;
\item
Higgs fields have a non-zero v.e.v.s, hence they cannot be
 superpartners of quarks and leptons since
this would induce a spontaneous violation of baryon and lepton
numbers;
\item
One needs at least two complex chiral Higgs multiplets to give
masses to Up and Down quarks.
\end{enumerate}

The latter is due to the form of a superpotential and chirality of
matter superfields. Indeed, the superpotential should be invariant
under $SU(3)\times SU(2)\times U(1)$ gauge group. If one looks at
the Yukawa interaction in the Standard Model, eq.(\ref{yuk}), one
finds that it is indeed $U(1)$ invariant since the sum of
hypercharges in each vertex equal zero. In the last term this is
achieved by taking  the conjugated Higgs doublet
$\tilde{H}=i\tau_2H^\dagger$ instead of $H$. However, in SUSY $H$
is a chiral superfield and hence a superpotential, which is
constructed out of  chiral fields, can contain only $H$ but
not $\tilde H$, which is an antichiral superfield.

Another reason for the second  Higgs doublet is related to chiral
anomalies. It is known that chiral anomalies spoil the gauge
invariance and, hence, the renormalizability of the theory. They
are canceled in the SM between quarks and leptons in each
generation.  However, if one introduces a chiral Higgs superfield,
it contains higgsinos, which are chiral fermions, and contain
anomalies.   To cancel them one has to add the second Higgs
doublet with the opposite hypercharge.

Therefore the Higgs sector in SUSY models is inevitably enlarged,
it contains an even number of doublets.

\vglue 0.4cm {\em Conclusion}: In SUSY models supersymmetry
associates {\em known} bosons with {\em new} fermi\-ons and {\em
known} fermi\-ons with {\em new} bosons.

\subsection{The field content}

Consider the particle content of the Minimal Supersymmetric
Standard Model \cite{MSSM}. According to the previous discussion
in the minimal version  we double the number of particles
(introducing a superpartner to each particle) and add another
Higgs doublet (with its superpartner). The particle content of the
MSSM then looks as follows~\cite{haber}:

\begin{center}
{\bf Particle Content of the MSSM} \\
\bigskip
\nopagebreak[4]
\renewcommand{\tabcolsep}{0.1cm}
\begin{tabular}{lllccc}
Superfield & \ \ \ \ \ \ \ Bosons & \ \ \ \ \ \ \ Fermions &
$SU_c(3)$& $SU_L(2)$ & $U_Y(1)$ \\ \hline \hline Gauge  &&&&& \\
${\bf G^a}$   & gluon \ \ \ \ \ \ \ \ \ \ \ \ \ \ \  $g^a$ &
gluino$ \ \ \ \ \ \ \ \ \ \ \ \ \tilde{g}^a$ & 8 & 0 & 0 \\ ${\bf
V^k}$ & Weak \ \ \ $W^k$ \ $(W^\pm, Z)$ & wino, zino \
$\tilde{w}^k$ \ $(\tilde{w}^\pm, \tilde{z})$ & 1 & 3& 0 \\ ${\bf
V'}$   & Hypercharge  \ \ \ $B\ (\gamma)$ & bino \ \ \ \ \ \ \ \ \
\ \ $\tilde{b}(\tilde{\gamma })$ & 1 & 1& 0 \\ \hline Matter &&&&
\\ $\begin{array}{c} {\bf L_i} \\ {\bf E_i}\end{array}$ & sleptons
\ $\left\{
\begin{array}{l} \tilde{L}_i=(\tilde{\nu},\tilde e)_L \\ \tilde{E}_i =\tilde
e_R \end{array} \right. $ & leptons \ $\left\{ \begin{array}{l}
L_i=(\nu,e)_L
\\ E_i=e_R \end{array} \right.$ & $\begin{array}{l} 1 \\ 1 \end{array} $  &
$\begin{array}{l} 2 \\ 1 \end{array} $ & $\begin{array}{r} -1 \\ 2
\end{array} $ \\ $\begin{array}{c} {\bf Q_i} \\ {\bf U_i} \\ {\bf D_i}
\end{array}$ & squarks \ $\left\{ \begin{array}{l}
\tilde{Q}_i=(\tilde{u},\tilde d)_L \\ \tilde{U}_i =\tilde u_R \\
\tilde{D}_i =\tilde d_R\end{array}  \right. $ & quarks \ $\left\{
\begin{array}{l} Q_i=(u,d)_L \\ U_i=u_R \\ D_i=d_R \end{array}
\right.$ & $\begin{array}{l} 3
\\ 3^* \\ 3^* \end{array} $  & $\begin{array}{l} 2 \\ 1 \\ 1 \end{array} $ &
$\begin{array}{r} 1/3 \\ -4/3 \\ 2/3 \end{array} $ \\ \hline
Higgs  &&&& \\ $\begin{array}{c} {\bf H_1} \\ {\bf
H_2}\end{array}$ & Higgses \ $\left\{
\begin{array}{l} H_1 \\ H_2 \end{array}  \right. $ & higgsinos \ $\left\{
 \begin{array}{l} \tilde{H}_1 \\ \tilde{H}_2 \end{array} \right.$ &
$\begin{array}{l} 1 \\ 1 \end{array} $  & $\begin{array}{l} 2 \\ 2
\end{array} $ &
$\begin{array}{r} -1 \\ 1
\end{array} $
 \\ \hline \hline
\end{tabular}
\end{center}
\vglue .5cm

\noindent where $a=1,2,...,8$ and $k=1,2,3$ are $SU(3)$ and
$SU(2)$ indices, respectively, and $i=1,2,3$ is the generation
index. Hereafter tilde denotes a superpartner of an ordinary particle.

Thus, the characteristic feature of any supersymmetric
generalization of the SM is the presence of superpartners.  If
supersymmetry is exact, superpartners of ordinary particles should
have the same masses and have to be observed. The absence of them
at modern energies is believed to be explained by the fact that
their masses are very heavy, that means that supersymmetry should
be broken. Hence,  if the energy of accelerators is
 high enough, the superpartners will be created.

The presence of an extra Higgs doublet in SUSY model is a novel
feature of the theory. In the MSSM  one has two doublets with the
quantum numbers (1,2,-1) and (1,2,1), respectively:
\begin{equation}
H_1=\left( \begin{array}{c} H^0_1 \\ H_1^- \end{array}  \right) =
\left( \begin{array}{c} v_1 +\frac{\displaystyle
S_1+iP_1}{\sqrt{2}} \\ H^-_1 \end{array}\right), \ \ \ H_2=\left(
\begin{array}{c} H^+_2 \\ H_2^0 \end{array}  \right) = \left(
\begin{array}{c} H^+_2 \\ v_2 +\frac{\displaystyle
S_2+iP_2}{\sqrt{2}}
\end{array} \right),
\end{equation}
where  $v_i$ are the vacuum expectation values of the neutral components.

Hence, one has 8=4+4=5+3 degrees of freedom. As in the case of the
SM, 3 degrees of freedom can be gauged away, and one is left with
5 physical states compared to 1 state in the SM.

Thus, in the MSSM, as actually in any two Higgs doublet model,
 one has  five  physical Higgs bosons: two CP-even neutral,
one CP-odd neutral  and two charged. We consider the mass eigenstates below.

\subsection{Largangian of the MSSM}

 The Lagrangian of the MSSM consists of two parts; the
first part is SUSY generalization of the Standard Model, while the
second one represents the SUSY breaking as mentioned above.
\begin{equation}
 {\cal L}={\cal L}_{SUSY}+{\cal L}_{Breaking},
\end{equation}
where
\begin{equation}
 {\cal L}_{SUSY}={\cal L}_{Gauge}+{\cal L}_{Yukawa}
\end{equation}
and
\begin{eqnarray}
{\cal L}_{Gauge} &= &\sum_{SU(3),SU(2),U(1)}^{}\frac{1}{4}
\left(\int d^2\theta \ Tr W^\alpha W_\alpha + \int d^2\bar\theta \
Tr \bar W^{\dot \alpha}\bar W_{\dot \alpha} \right) \nonumber \\
&& +\sum_{Matter}^{}\int d^2\theta d^2 \bar\theta \
\Phi^\dagger_ie^{\displaystyle g_3\hat V_3 + g_2\hat V_2 + g_1\hat
V_1}\Phi_i ,  \\ {\cal L}_{Yukawa}&= &\int d^2\theta \ ({\cal
W}_R+{\cal W}_{NR}) + h.c .
\end{eqnarray}
The index $R$ in a superpotential refers to the so-called
$R$-parity  which adjusts a "$+$" charge to all the ordinary
particles and a "$-$" charge to their superpartners \cite{r-parity}.
The first part of ${\cal W}$ is R-symmetric
\begin{equation}
W_{R} = \epsilon_{ij}(y^U_{ab}Q_a^j U^c_bH_2^i +
y^D_{ab}Q_a^jD^c_bH_1^i
       +  y^L_{ab}L_a^jE^c_bH_1^i + \mu H_1^iH_2^j), \label{R}
\end{equation}
where $i,j=1,2,3$ are $SU(2)$ and $a,b=1,2,3$ are the generation
indices; colour indices are suppressed. This part of the
Largangian almost exactly repeats that of the SM except that the
fields are now the superfields rather than the ordinary fields of
the SM. The only difference is the last term which describes the
Higgs mixing. It is absent in the SM since we have only one Higgs
field there.

The second part is R-nonsymmetric
\begin{eqnarray}
W_{NR} &=& \epsilon_{ij}(\lambda^L_{abd}L_a^i L_b^jE_d^c +
\lambda^{L\prime}_{abd}L_a^iQ_b^jD_d^c +\mu'_aL^i_aH_2^j)
\nonumber\\ & + &  \lambda^B_{abd}U_a^cD_b^cD_d^c. \label{NR}
\end{eqnarray}
These terms are absent in the SM. The reason is very simple: one
can not replace the superfields in eq.(\ref{NR}) by the ordinary
fields like in eq.(\ref{R}) because of the Lorentz invariance.
These terms have the other property, they violate either lepton
(the first line in eq.(\ref{NR})) or baryon number (the second
line). Since both effects are not observed in Nature, these terms
must be suppressed or be excluded.  In the minimal version of the
MSSM these terms are not included, they are forbidden by R-parity
conservation \cite{r-con}.

\subsection{Soft SUSY breaking}

To introduce supersymmetry breaking as required by the absence of
superpartners at modern energies, one has to be careful not to
spoil the cancellation of divergencies which allows to solve the
hierarchy problem. This is achieved by spontaneous breaking of
SUSY in the same way as spontaneous electroweak symmetry breaking.
One introduces the field whose vacuum expectation value breaks
supersymmetry. However, due to a special character of SUSY, this
should be a superfield, whose auxiliary $F$ or $D$ components
acquire non-zero v.e.v.'s.  This leads to  appearance of the
so-called soft breaking terms. In the simplest version the soft
supersymmetry breaking terms are \cite{breaking}:
\begin{eqnarray}
-{\cal L}_{Breaking} & = & m_0^2\sum_{i}^{}|\varphi_i|^2+\left(
\frac 12 m_{1/2}\sum_{\alpha}^{} \tilde \lambda_\alpha\tilde
\lambda_\alpha\right.  \label{soft}\\
 & + & \left. A[y^U_{ab}\tilde Q_a\tilde U^c_bH_2+y^D_{ab}\tilde Q_a
 \tilde D^c_bH_1+ y^L_{ab}\tilde L_a\tilde E^c_bH_1]
 +  B[\mu H_1H_2] +h.c.\right) , \nonumber
\end{eqnarray}
where we have suppressed $SU(2)$ indices.  Here $\varphi_i$  are
all scalar fields, $\tilde \lambda_\alpha $  are the gaugino
fields, $\tilde Q, \tilde U, \tilde D$ and $\tilde L, \tilde E$ are
the squark and slepton  fields, respectively, and
$H_{1,2}$  are the SU(2) doublet Higgs fields.

These terms are obtained via supergravity mechanism and are
usually introduced at the GUT scale. We have assumed in
eq.(\ref{soft}) the so-called universality of the soft terms,
namely, we put all the spin 0 particle masses to be equal to the
universal value $m_0$, all the spin 1/2 particle (gaugino) masses
to be equal to $m_{1/2}$ and all the cubic and quadratic terms,
proportional to $A$ and $B$, to repeat the structure of the Yukawa
superpotential (\ref{R}). This is an additional requirement motivated by the
supergravity mechanism of SUSY breaking as mentioned earlier
\cite{SUGRA}.

Universality is not a necessary requirement and one may consider
non-universal soft terms as well. However, it will not change the
qualitative picture presented below, so for simplicity in what follows we consider
 the universal boundary conditions.

It should be noted that supergravity induced universality of the
soft terms is more likely to be valid at the Planck scale, rather
than at the GUT one. This is because a natural scale for gravity
is $M_{Planck}$, while $M_{GUT}$ is the scale for the gauge
interactions. However, due to a small difference between these two
scales, it is usually ignored in the first approximation resulting
in minor uncertainties in the low-energy predictions
\cite{polonsky}.

The soft terms explicitly break supersymmetry. As will be shown
later they lead to the mass spectrum of superpartners different
from that of the ordinary particles. Remind that the masses of
quarks and leptons remain zero until $SU(2)$ invariance is
spontaneously broken.

\subsection{Masses}

With given values of $m_0,m_{1/2},\mu,Y_t,Y_b,Y_\tau, A$, and $B$
at the GUT scale, one can solve the corresponding RG equations
thus linking the values at the GUT and electroweak scales.
Substituting these parameters into the mass matrices one can
predict the mass spectrum of superpartners \cite{MSSM, spectrum}.

\subsubsection{Gaugino-higgsino mass terms}

The mass matrix for the gauginos, the superpartners of the gauge
bosons, and for higgsinos, the superpartners of the Higgs bosons,
is non-diagonal, thus leading to their mixing. The mass terms look
like
\begin{equation}
{\cal L}_{Gaugino-Higgsono}=
 -\frac{1}{2}M_3\bar{\lambda}_a\lambda_a
 -\frac{1}{2}\bar{\chi}M^{(0)}\chi -(\bar{\psi}M^{(c)}\psi + h.c.),
\end{equation}
where $\lambda_a , a=1,2,\ldots ,8,$ are the Majorana gluino
fields and
\begin{equation}
\chi = \left(\begin{array}{c}\tilde{B}^0 \\ \tilde{W}^3 \\
\tilde{H}^0_1 \\ \tilde{H}^0_2
\end{array}\right), \ \ \ \psi = \left( \begin{array}{c}
\tilde{W}^{+} \\ \tilde{H}^{+}
\end{array}\right)
\end{equation}
are, respectively, the Majorana neutralino and Dirac chargino
fields. The neutralino mass matrix is:
\begin{equation}
M^{(0)}=\left(
\begin{array}{cccc}
M_1 & 0 & -M_Z\cos\beta \sin_W & M_Z\sin\beta \sin_W \\ 0 & M_2 &
M_Z\cos\beta \cos_W   & -M_Z\sin\beta \cos_W  \\ -M_Z\cos\beta
\sin_W & M_Z\cos\beta \cos_W  & 0 & -\mu \\ M_Z\sin\beta \sin_W &
-M_Z\sin\beta \cos_W  & -\mu & 0
\end{array} \right).\label{neut}
\end{equation}
The physical neutralino masses  $M_{\tilde{\chi}_i^0}$ are
obtained as eigenvalues of this matrix after diagonalization. For
charginos one has:
\begin{equation}
M^{(c)}=\left(
\begin{array}{cc}
M_2 & \sqrt{2}M_W\sin\beta \\ \sqrt{2}M_W\cos\beta & \mu
\end{array} \right).\label{char}
\end{equation}
This matrix has two chargino eigenstates
$\tilde{\chi}_{1,2}^{\pm}$ with mass eigenvalues
\begin{equation}
M^2_{1,2}=\frac{1}{2}\left[M^2_2+\mu^2+2M^2_W \mp
\sqrt{(M^2_2-\mu^2)^2+4M^4_W\cos^22\beta
+4M^2_W(M^2_2+\mu^2+2M_2\mu \sin 2\beta )}\right].
\end{equation}

\subsubsection{Squark and slepton masses}

The non-negligible Yukawa couplings cause a mixing between the
electroweak eigenstates and the mass eigenstates of the third
generation particles.  The mixing matrices for the
$\tilde{m}^{2}_t,\tilde{m}^{2}_b$ and $\tilde{m}^{2}_\tau$ are:
\begin{equation} \label{stopmat}
\left(\begin{array}{cc} \tilde m_{tL}^2& m_t(A_t-\mu\cot \beta )
\\ m_t(A_t-\mu\cot \beta ) & \tilde m_{tR}^2 \end{array}  \right),
\nonumber
\end{equation}
\begin{equation} \label{sbotmat}
\left(\begin{array}{cc} \tilde  m_{bL}^2& m_b(A_b-\mu\tan \beta )
\\ m_b(A_b-\mu\tan \beta ) & \tilde  m_{bR}^2 \end{array}
\right), \nonumber
\end{equation}
\begin{equation} \label{staumat} \left(\begin{array}{cc}
\tilde  m_{\tau L}^2& m_{\tau}(A_{\tau}-\mu\tan \beta ) \\
m_{\tau}(A_{\tau}-\mu\tan \beta ) & \tilde m_{\tau R}^2
\end{array}  \right)               \nonumber
\end{equation}
with
\begin{eqnarray*}
  \tilde m_{tL}^2&=&\tilde{m}_Q^2+m_t^2+\frac{1}{6}(4M_W^2-M_Z^2)\cos
  2\beta ,\\
  \tilde m_{tR}^2&=&\tilde{m}_U^2+m_t^2-\frac{2}{3}(M_W^2-M_Z^2)\cos
  2\beta ,\\
  \tilde m_{bL}^2&=&\tilde{m}_Q^2+m_b^2-\frac{1}{6}(2M_W^2+M_Z^2)\cos
  2\beta ,\\
  \tilde m_{bR}^2&=&\tilde{m}_D^2+m_b^2+\frac{1}{3}(M_W^2-M_Z^2)\cos
  2\beta ,\\
 \tilde m_{\tau L}^2&=&\tilde{m}_L^2+m_{\tau}^2-\frac{1}{2}(2M_W^2-M_Z^2)\cos
2\beta ,\\ \tilde m_{\tau
R}^2&=&\tilde{m}_E^2+m_{\tau}^2+(M_W^2-M_Z^2)\cos
  2\beta
\end{eqnarray*}
and the  mass eigenstates are  the eigenvalues of these mass
matrices.

\subsubsection{The Higgs potential}

As has been already  mentioned, the Higgs potential in MSSM is
totally defined by superpotential (and the soft terms). Due to the
structure of ${\cal W}$ the Higgs self-interaction is given by the
$D$-terms, while the $F$-terms contribute only to the mass matrix.
The tree level potential is:
\begin{eqnarray}
V_{tree}(H_1,H_2)&=&m^2_1|H_1|^2+m^2_2|H_2|^2-m^2_3(H_1H_2+h.c.)
\label{Higpot} \nonumber\\ &+&
\frac{g^2+g^{'2}}{8}(|H_1|^2-|H_2|^2)^2 +
\frac{g^2}{2}|H_1^+H_2|^2,
\end{eqnarray}
where $m_1^2=m^2_{H_1}+\mu^2, m_2^2=m^2_{H_2}+\mu^2$. At the GUT scale
$m_1^2=m^2_2=m_0^2+\mu^2_0, \ m^2_3=-B\mu_0$. Notice, that the Higgs
self-interaction coupling in eq.(\ref{Higpot}) is  fixed and is defined by
the gauge interactions as opposite to the SM.

The potential (\ref{Higpot}), in accordance with supersymmetry, is
positively definite and stable. It has no non-trivial minimum
different from zero.  Indeed, let us write the minimization condition for  the potential
(\ref{Higpot})
\begin{eqnarray} \frac 12\frac{\delta V}{\delta
H_1}&=&m_1^2v_1 -m^2_3v_2+ \frac{g^2+g'^2}4(v_1^2-v_2^2)v_1=0,
\label{min1}
\\ \frac 12\frac{\delta V}{\delta H_2}&=&m_2^2v_2-m^2_3v_1+ \frac{
g^2+g'^2}4(v_1^2-v_2^2)v_2=0, \label{min2}
\end{eqnarray}
where we have  introduced the notation $$<H_1>\equiv v_1= v
\cos\beta , \ \  <H_2>\equiv v_2= v \sin\beta, \ \ \ \ v^2=
v_1^2+v_2^2,\ \ \  \tan\beta \equiv \frac{v_2}{v_1}.$$ Solution of
eqs.(\ref{min1}),(\ref{min2}) can be expressed in terms of $v^2$
and $\sin\beta$:
\begin{equation} v^2=\frac{\displaystyle  4(m^2_1-m^2_2\tan^2 \beta
)}{\displaystyle (g^2+ g'^2)(\tan^2\beta -1)},\ \ \ \sin2\beta
=\frac{\displaystyle 2m^2_3}{\displaystyle m^2_1+m^2_2}.
\label{min}
\end{equation}
One can easily see from eq.(\ref{min}) that if
$m_1^2=m_2^2=m_0^2+\mu_0^2$, $v^2$ happens to be negative, i.e.
the minimum does not exist.  In fact, real positive solutions to
eqs.(\ref{min1}),(\ref{min2}) exist only if the following
conditions are satisfied \cite{haber}:
\begin{equation}
m_1^2+m_2^2 > 2 m_3^2, \ \ \ \  m_1^2m_2^2 < m_3^4 , \label{cond}
\end{equation}
which is not the case at the GUT scale.
This means that spontaneous breaking of the
$SU(2)$  gauge invariance, which is needed in the SM to give
masses for all the particles, does not take place in the MSSM.

This strong statement is valid, however, only at the GUT scale.
Indeed, going down with energy the parameters of the potential
(\ref{Higpot}) are renormalized.  They become the ``running''
parameters with the energy scale dependence given by the RG
equations. The running of the parameters leads to a remarkable
phenomenon known as a {\em radiative spontaneous symmetry
breaking} which we discuss below.

Provided conditions (\ref{cond}) are satisfied the
mass matrices at the tree level are \\
CP-odd components
 $P_1$ and $P_2$ :
\begin{equation}
{\cal M}^{odd} = \left.\frac{\partial^2 V}{\partial P_i \partial P_j}
\right |_{H_i=v_i}
= \left( \begin{array}{cc}  \tan\beta &1 \\1& \cot\beta \end{array}\right) m_3^2,
\end{equation}
 CP-even neutral components $S_1$ and $S_2$:
\begin{equation}
{\cal M}^{even} = \left.\frac{\partial^2 V}{\partial S_i \partial S_j}
\right|_{H_i=v_i}
= \left( \begin{array}{cc}  \tan\beta & -1 \\-1& \cot\beta \end{array}\right) m_3^2
+\left( \begin{array}{cc}  \cot\beta & -1 \\-1& \tan\beta \end{array}\right)
M_Z\cos\beta \sin\beta,
\end{equation}
Charged components $H^-$ and $H^+$:
\begin{equation}
{\cal M}^{charged} =\left.\frac{\partial^2 V}{\partial H^+_i \partial H^-_j}
\right|_{H_i=v_i}
= \left( \begin{array}{cc}  \tan\beta &1 \\1& \cot\beta \end{array}\right)
 (m_3^2+M_W\cos\beta\sin\beta).
\end{equation}
Diagonalising the mass matrices one gets the mass eigenstates
\cite{haber}: $$\begin{array}{l} \left\{
\begin{array}{lllr} G^0 \ &=& -\cos\beta P_1+\sin \beta P_2 , & \
\ \ Goldstone \ boson \ \to Z_0,
\\ A \ &=& \sin\beta P_1+\cos \beta P_2 , & \ \ \ \ \ \ \ \ Neutral \ CP=-1 \
Higgs, \end{array}\right.\\  \\ \left\{ \begin{array}{lllr} G^+
&=& -\cos\beta (H^-_1)^*+\sin \beta H^+_2 , &\ \ Goldstone \ boson
\ \to W^+, \\ H^+ &=& \sin\beta (H^-_1)^*+\cos \beta H^+_2 , &\ \
\ Charged \ Higgs,
\end{array}\right.\\ \\
\left\{ \begin{array}{lllr} h \ &=& -\sin\alpha S_1+\cos\alpha S_2
, & \ \ \ \ \ \ \ SM \ Higgs \ boson \ CP=1, \\ H \ &=& \cos\alpha
S_1+\sin\alpha S_2 , & \ \ \ \ \ \ \ Extra \ heavy \ Higgs \ boson
,
\end{array}\right.
\end{array}$$
where the mixing angle $\alpha $ is given by $$
\tan 2\alpha = -\tan 2\beta
\left(\frac{m^2_A+M^2_Z}{m^2_A-M^2_Z}\right).$$
The physical Higgs
bosons acquire the following masses \cite{MSSM}:
\begin{eqnarray}
\mbox{CP-odd neutral Higgs} \ \ A: && \ \ \ \ \ \ \ \ \ \ \ \
m^2_A = m^2_1+m^2_2, \nonumber \\ \mbox{Charge Higgses} \ \
H^{\pm}: && \ \ \ \ \ \
 \ \ \ \ \ m^2_{H^{\pm}}=m^2_A+M^2_W , \\
\mbox{CP-even neutral Higgses} \ \ H,h: && \nonumber
\end{eqnarray}
\begin{equation} m^2_{H,h}=
\frac{1}{2}\left[m^2_A+M^2_Z \pm
\sqrt{(m^2_A+M_Z^2)^2-4m^2_AM_Z^2\cos^22\beta}\right],
\end{equation}
where as usual $$ M^2_W=\frac{g^2}{2}v^2, \ \
M^2_Z=\frac{g^2+g'^2}{2}v^2 .$$
This leads to the once celebrated SUSY mass relations:
$$\begin{array}{c} m_{H^{\pm}} \geq M_W, \\[0.2cm] m_h \leq m_A
\leq M_H, \\[0.3cm] m_h \leq M_Z |\cos 2\beta| \leq M_Z ,\\[0.3cm]
m_h^2+m_H^2=m_A^2+M_Z^2.
\end{array}$$

Thus, the lightest neutral Higgs boson happens to be lighter than
$Z$ boson, that clearly distinguishes it from the SM one. Though
we do not know the mass of the Higgs boson in the SM, there are
several indirect constraints leading to the lower boundary of $m_h^{SM}
\geq 135 $ GeV~\cite{bound1}. After including the radiative
corrections  the mass of the lightest Higgs boson in the MSSM, $m_h$,  increases.
We consider it in more detail below.

\subsection{RG flow in the MSSM}

If one compares the RG flow in the SM and the MSSM, one finds
additional contributions from superpartners to the RG equations.

\begin{figure}[tb]
\vspace{-2cm}
\begin{center}
  \leavevmode
  \epsfxsize=15cm
 \epsffile{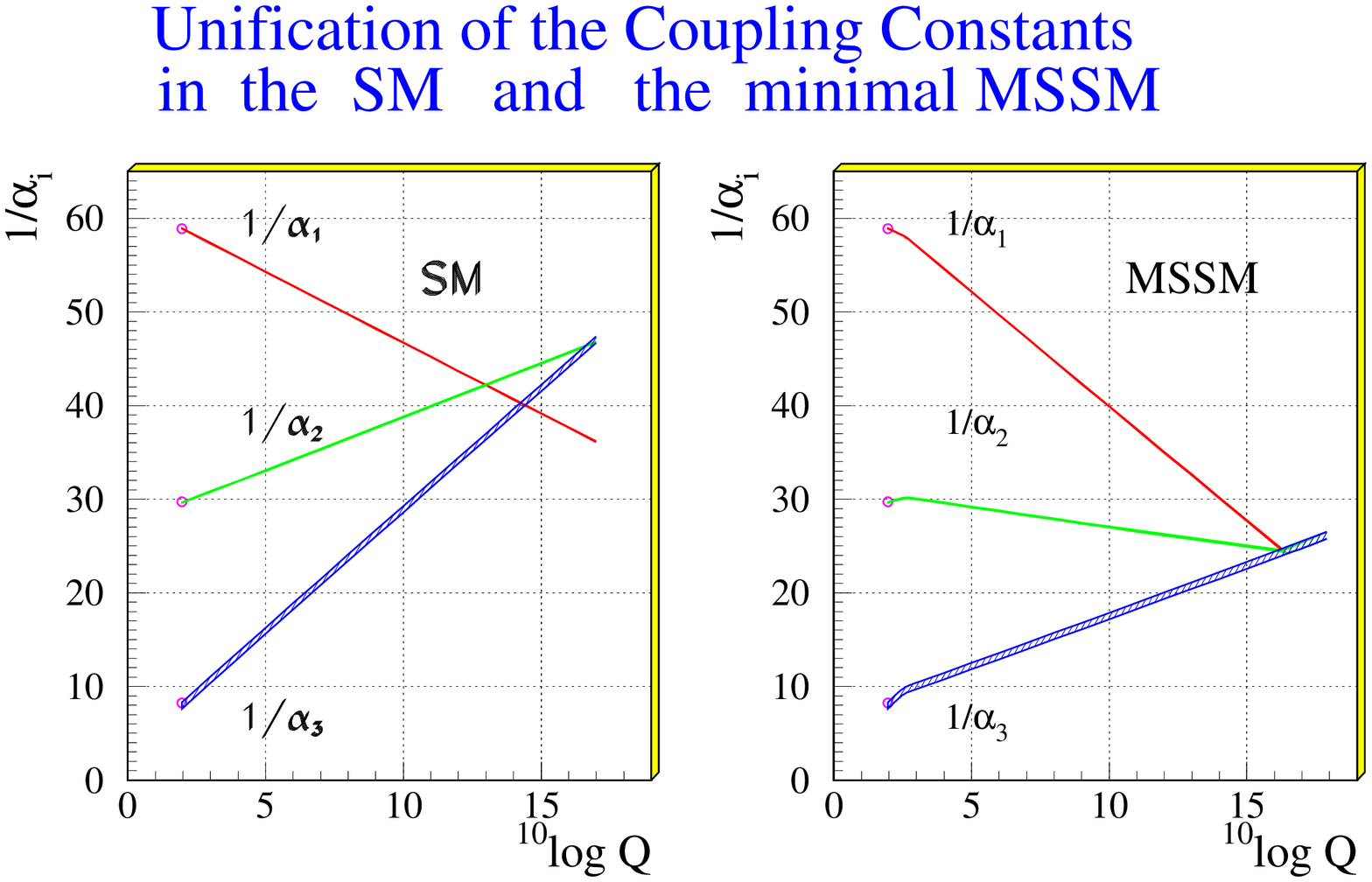}
\end{center}
\caption{Evolution of the inverse of the three coupling constants
in the Standard Model (left) and in the supersymmetric extension
of the SM (MSSM) (right). Only in the latter case unification is
obtained. The SUSY particles are assumed to contribute only above
the effective SUSY scale $M_{SUSY}$ of about  1 TeV, which causes
the change in slope in the evolution of couplings. The thickness
of the lines represents the error in the coupling constants
\cite{ABF}.} \label{unif}
\end{figure}

Consider the  gauge couplings.  In the SM the RG flow is given by
eqs.(\ref{alpha}). We have mentioned already in Sec.1 that it
offers the unification  pattern supporting the GUT hypothesis.
However, if one looks at the curves more attentively,  one finds
that the situation is not that good. Indeed, let us consider the
solution to the RG equations in more detail. The result is
demonstrated in the left part of Fig.\ref{unif}, which shows the
evolution of an inverse of the couplings as function of a
logarithm of energy~\cite{ABF}. In this presentation the evolution
becomes a straight line in first order. The second order
corrections are small and do not cause any visible deviation from
a straight line. Fig.\ref{unif} clearly demonstrates that within
the SM the coupling constants unification at a single point is
impossible. It is excluded by more than 8 standard
deviations~\cite{ABF,LL,EKN}. This result means that the
unification can only be obtained if new physics enters between the
electroweak and the Planck scales.

 In the MSSM the slopes of the RG evolution curves are modified. The
coefficients $b_i$ in eq.(\ref{alpha}) now are :
\begin{equation}
b_i=\left( \begin{array}{r} b_1 \\ b_2 \\b_3 \end{array} \right)
   =
\left( \begin{array}{r}          0     \\
                                -6     \\
                                -9      \end{array} \right) +N_{Fam}
\left( \begin{array}{r}          2     \\
                                 2     \\
                                 2      \end{array} \right) +N_{Higgs}
\left( \begin{array}{r}          3/10  \\
                                 1/2   \\
                                 0      \end{array} \right)     ,
\label{susy1}
\end{equation}
where use  $N_{Fam}=3$ and $N_{Higgs}=2$, which corresponds to the
MSSM.
%
%

 It turns out that within the SUSY model
perfect unification can be obtained if the SUSY  masses are of the
order of 1 TeV. This is shown on the right part of Fig.\ref{unif};
the SUSY particles are assumed to contribute effectively to the
running of the coupling constants only for energies above the
typical SUSY mass scale, which causes the change in the slope of
the lines near 1 TeV. From the fit requiring unification one finds
for the breakpoint $M_{SUSY}$ and the unification point
$M_{GUT}$~\cite{ABF}:
\begin{eqnarray}
M_{SUSY} &= & 10^{3.4\pm 0.9\pm 0.4} \ GeV , \nonumber\\
M_{GUT} &=& 10^{15.8\pm 0.3\pm 0.1} \ GeV , \label{Msusy} \\
\alpha^{-1}_{GUT} &= & (26.3 \pm 1.9 \pm 1.0).\nonumber
\end{eqnarray}
The first error originates from the uncertainty in the coupling
constant, while the second one is due to the uncertainty in the
mass splittings between the SUSY particles. For SUSY models, the
dimensional reduction $\overline{DR}$ scheme  is used \cite{DR}.

This unification of the gauge couplings was considered as the
first ``evidence'' for supersymmetry, especially since $M_{SUSY}$
was found in the range preferred by the fine-tuning arguments.

It should be noted, that the unification of the three curves at a
single point is not that trivial as it may seem from the existence
of  three free parameters ($ M_{SUSY}, M_{GUT}$ and
$\alpha_{GUT}$).  The reason is simple: when introducing new
particles one influences all three curves simultaneously, thus
giving rise to strong correlations between the slopes of the three
lines. For example, adding new generations and/or new Higgs
doublets never yield unification.

 \section{Renormalization of Softly Broken SUSY Theories}
\setcounter{equation} 0

To find the RG flow for the soft terms one has to know how they
are renormalized. Remarkably  that the renormalizations in softly
broken SUSY theories follow a simple pattern which is completely
defined by an unbroken theory \cite{Yamada, JJ,AKK, GR}.

The main idea is that a softly broken supersymmetric gauge theory
can be considered as a rigid SUSY theory imbeded into  external
space-time independent  superfield, so that all  couplings and
masses become external superfields. The crucial statement is that
the {\em singular} part of effective action depends on external
superfield, but not on its derivatives, so that one can calculate
it when the external field is a constant, i.e. in a rigid theory
\cite{Van}. This approach to a softly broken sypersymmetric theory
allows one to use remarkable mathematical properties of $N=1$ SUSY
theories such as non-renormalization theorems, cancellation of
quadratic divergences, etc. The renormalization procedure in a
softly broken SUSY gauge theory can be performed in the following
way \cite{AKK}:

 One takes  renormalization constants of a rigid theory,
calculated in some massless scheme, substitutes instead of the
rigid couplings (gauge and Yukawa) their modified expressions,
which depend on a Grassmannian variable, and expand over this
variable.

This gives   renormalization constants for the soft terms.
Differentiating them with respect to a scale one can find
corresponding renormalization group equations.

In fact as it has been shown in \cite{Kaz} this procedure works at
all stages. One can make the above mentioned substitution on the
level of the renormalization constants, RG equations, solutions to
these equations, approximate solutions, fixed points, finiteness
conditions, etc. Expanding then over a Grassmannian variable one
obtains  corresponding expressions for the soft terms.

We demonstrate now how this procedure works in the MSSM. Using notation
introduced above the modified couplings in the MSSM are ($\eta=\theta^2,
\bar \eta = \bar \theta^2$)
\begin{eqnarray}
\tilde{{\mbox a}}_i&=&{\mbox a}_i(1+M_i \eta+\bar M_i
\bar{\eta}+(M_i\bar M_i+\Sigma_{\alpha_i})
 \eta \bar{\eta}), \label{ga}\\
\tilde{Y}_k&=&Y_k(1-A_k \eta -\bar A_k \bar{\eta}+ (A_k\bar A_k+\Sigma_k)
\eta \bar \eta),
 \label{ya}
\end{eqnarray}
where $M_i$ are the gaugino masses, $A_k$ are the trilinear scalar
couplings, $\Sigma_k$ are the certain combinations of soft squark
and slepton masses entering the Yukawa vertex and $\Sigma_{\alpha_i}$ are
the SUSY ghost soft terms
 $$ \Sigma_t= \tilde{m}_{Q3}^2 + \tilde{m}_{U3}^2  +
m^2_{H2},  \ \ \Sigma_b= \tilde{m}_{Q3}^2 + \tilde{m}_{D3}^2  +
m^2_{H1},  \ \ \Sigma_\tau= \tilde{m}_{L3}^2 + \tilde{m}_{E3}^2 +
m^2_{H1},  \ \  \Sigma_{\alpha_i}=M_i^2+\tilde{m}^2_{gh_i}$$ and
$\tilde{m}^2_{gh}$ is the soft scalar ghost mass, which is
eliminated by solving the RG equation.
 In one-loop order  $\tilde{m}^2_{gh}=0$.

  To get now the
RG equations for the soft terms one just has to take the
corresponding RG equations for the rigid couplings and perform the
Grassmannian expansion. The one-loop RG equations for the MSSM
couplings are \cite{Ibanez}:
\begin{eqnarray}
\frac{d{\mbox a}_i}{dt}&=&b_i {\mbox a}_i^2, \nonumber\\ \frac{dY_U}{dt}
& = & -Y_L\left(\frac{16}{3}{\mbox a}_3 + 3{\mbox a}_2 +
\frac{13}{15}{\mbox a}_1-6Y_U-Y_D\right) , \nonumber \\
\frac{dY_D}{dt} & = & -Y_D\left(\frac{16}{3}{\mbox a}_3 + 3{\mbox a}_2 +
\frac{7}{15}{\mbox a}_1-Y_U-6Y_D-Y_L\right), \nonumber
\\ \frac{dY_L}{dt} & = &- Y_L\left( 3{\mbox a}_2 +
  \frac{9}{5}{\mbox a}_1-3Y_D-4Y_L\right). \label{eq}
\end{eqnarray}
Performing the Grassmannian expansion one finds:
\begin{eqnarray}
\frac{dM_i}{dt} & = & b_i {\mbox a}_iM_i . \nonumber\\
\frac{dA_U}{dt} & = & \frac{16}{3}{\mbox a}_3 M_3 + 3{\mbox a}_2
M_2 + \frac{13}{15}{\mbox a}_1 M_1+6Y_UA_U+Y_DA_D, \nonumber
\\
 \frac{dA_D}{dt} & = & \frac{16}{3}{\mbox a}_3 M_3 +
3{\mbox a}_2 M_2 + \frac{7}{15}{\mbox a}_1 M_1+6Y_DA_D+Y_UA_U+Y_LA_L,
\nonumber \\ \frac{dA_L}{dt} & = &  3{\mbox a}_2 M_2 +
\frac{9}{5}{\mbox a}_1 M_1+3Y_DA_D+4Y_LA_L, \nonumber
\\ \frac{dB}{dt} & = & 3{\mbox a}_2 M_2 +
\frac{3}{5}{\mbox a}_1 M_1+3Y_UA_U+3Y_DA_D+Y_LA_L. \nonumber
\\ &&\nonumber \\
 \frac{d\tilde{m}^2_Q}{dt} & =&- \left[
(\frac{16}{3}{\mbox a}_3M^2_3 + 3{\mbox a}_2M^2_2 +
\frac{1}{15}{\mbox a}_1M^2_1)
-Y_U(\tilde{m}^2_Q+\tilde{m}^2_U+m^2_{H_2}+A^2_U)\right. \nonumber
\\
 & &\left. -Y_D(\tilde{m}^2_Q+\tilde{m}^2_D+m^2_{H_1}+A^2_D)\right], \nonumber\\
\frac{d\tilde{m}^2_U}{dt} & = &- \left[(\frac{16}{3}{\mbox a}_3M^2_3
+\frac{16}{15}{\mbox a}_1M^2_1)
-2Y_U(\tilde{m}^2_Q+\tilde{m}^2_U+m^2_{H_2}+A^2_U)\right] ,
\nonumber\\ \frac{d\tilde{m}^2_D}{dt} & =
 &- \left[(\frac{16}{3}{\mbox a}_3M^2_3+ \frac{4}{15}{\mbox a}_1M^2_1)
-2Y_D(\tilde{m}^2_Q+\tilde{m}^2_D+m^2_{H_1}+A^2_D)\right],
\nonumber\\
  \frac{d\tilde{m}^2_L}{dt} & = &
-\left[3(
 {\mbox a}_2M^2_2 + \frac{1}{5}{\mbox a}_1M^2_1)
-Y_L(\tilde{m}^2_L+\tilde{m}^2_E+m^2_{H_1}+A^2_L)\right],
\nonumber
\\ \frac{d\tilde{m}^2_E}{dt} & = &-\left[ (
 \frac{12}{5}{\mbox a}_1M^2_1)-2Y_L(
\tilde{m}^2_L+\tilde{m}^2_E+m^2_{H_1}+A^2_L)\right], \nonumber
\\
\frac{d\mu^2}{dt}&=&-\mu^2\left[3({\mbox a}_2+
\frac{1}{5}{\mbox a}_1)-(3Y_U+3Y_D+Y_L)\right],\label{eq2} \\
\frac{dm^2_{H_1}}{dt} & = & -\left[3({\mbox a}_2M^2_2
+\frac{1}{5}{\mbox a}_1M^2_1)
-3Y_D(\tilde{m}^2_Q+\tilde{m}^2_D+m^2_{H_1}+A^2_D)\right.
\nonumber
\\ &&\left. - Y_L(\tilde{m}^2_L+\tilde{m}^2_E+m^2_{H_1}+A^2_L)\right] ,
\nonumber\\ \frac{dm^2_{H_2}}{dt} & = &-\left[ 3({\mbox a}_2M^2_2
+\frac{1}{5}{\mbox a}_1M^2_1)
-3Y_U(\tilde{m}^2_Q+\tilde{m}^2_U+m^2_{H_2}+A^2_U)\right],
\nonumber
\end{eqnarray}
where we have already substituted the solution
$\tilde{m}_{gh}^2=0$ in the one-loop order. (Note that to get the
RG equation for the individual squark and slepton masses one needs
to know the anomalous dimensions for the corresponding fields.)

\section {RG Flow for the Soft Terms}
\setcounter{equation} 0

\begin{figure}[hbt]
\begin{flushleft}
\leavevmode \epsfxsize=7cm \epsffile{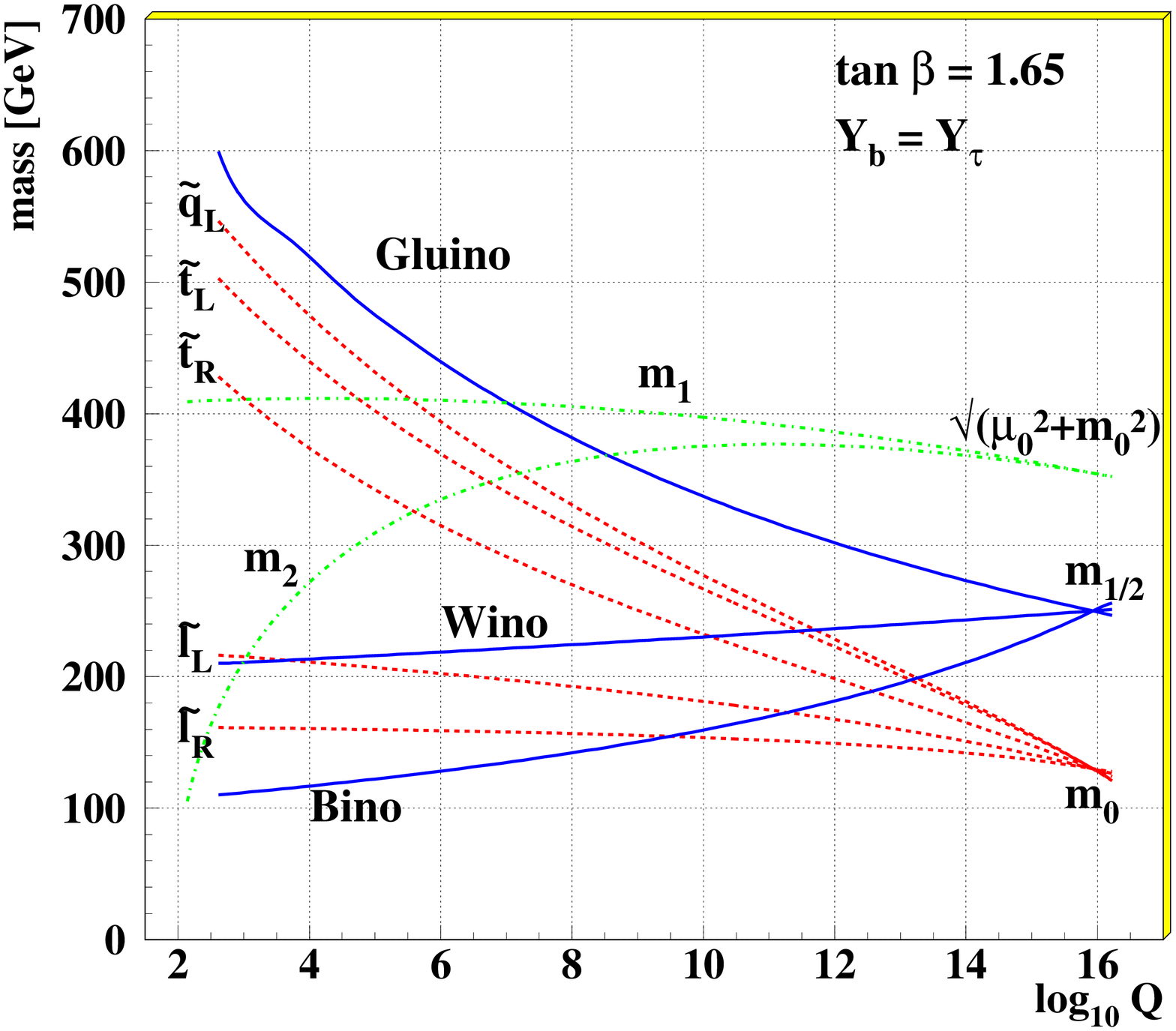} \end{flushleft}
\vspace{-7.5cm}\hspace{8cm}
 \leavevmode
       \epsfxsize=7cm
\epsffile{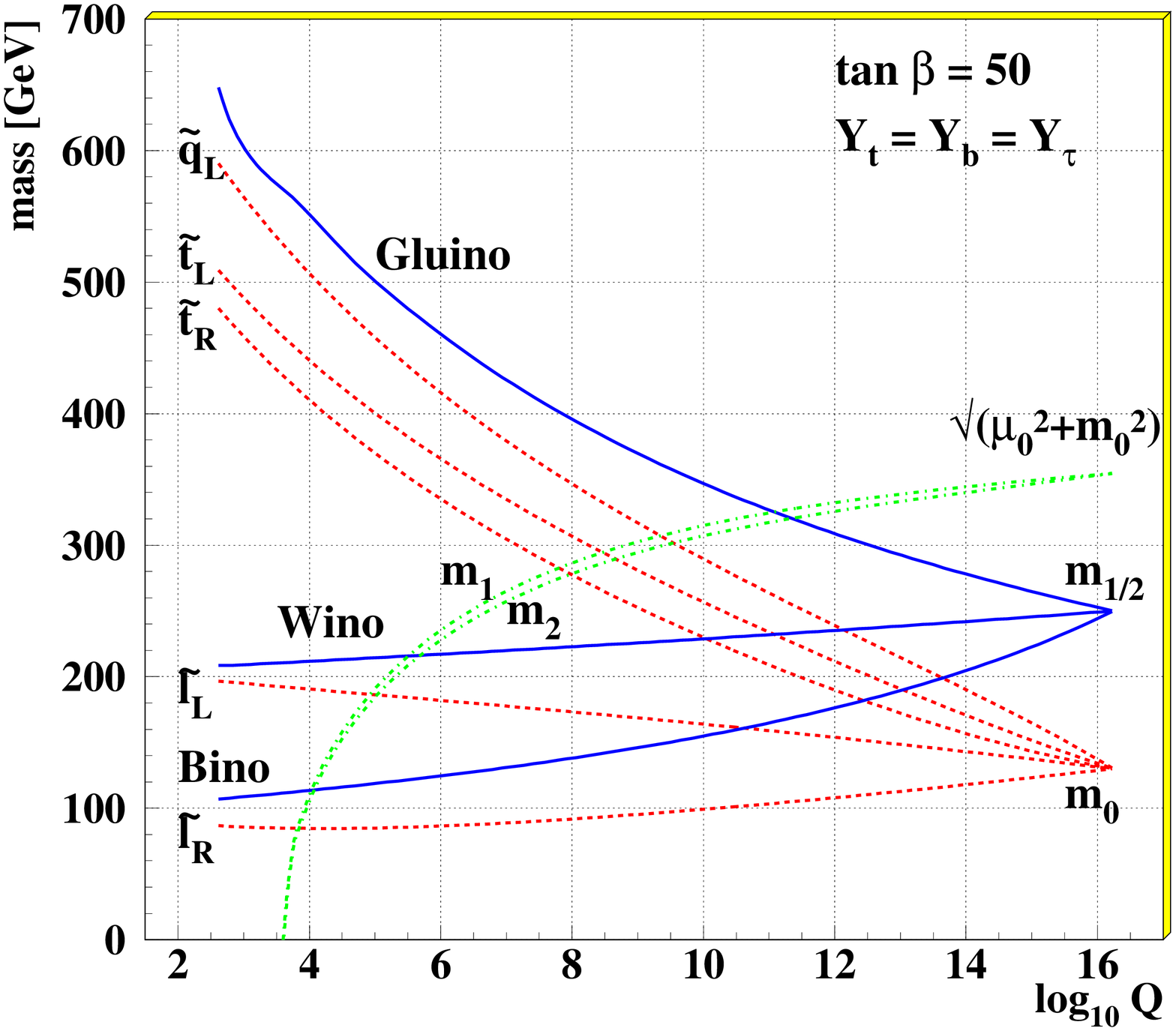}
\caption{An example of evolution of sparticle masses and soft
supersymmetry breaking parameters $m_1^2=m_{H_1}^2+\mu^2$ and
$m_2^2=m_{H_2}^2+\mu^2$.} \label{16}
\end{figure}

Having all the RG equations, one can find now the RG flow for the
soft terms. To see what happens at lower scales one has to run the
RG equations for the mass parameters from GUT to the EW scale. Let
us take some initial values of the soft masses at the GUT scale in
the interval between $10^2\div 10^3$ GeV consistent with SUSY
scale suggested by unification of the gauge couplings
(\ref{Msusy}). This leads to the following RG flow of the soft
terms shown in Fig.\ref{16}.~\cite{spectrum} (note that we perform
the running of soft paramerters in the opposite direction, from
GUT to EW scale)
%
%

One should mention the following general features common to any
choice of initial conditions:

 i) The gaugino masses follow the running of the gauge couplings
 and split at low energies. The gluino mass is running faster
 than the others  and is usually the heaviest due to the strong interaction.

 ii) The squark and slepton masses also split at low energies, the
 stops (and sbottoms) being the lightest due to relatively big Yukawa couplings
 of the third generation.

 iii) The Higgs masses (or at least one of them) are running down
 very quickly and may even become negative.

 To calculate the masses one has also to take into account the
mixing between various states (see eqs.(\ref{neut},\ref{char},
\ref{stopmat}-\ref{staumat}).

\subsection{Radiative Electroweak symmetry breaking}

The running of the Higgs masses leads to the phenomenon known as a
{\em radiative electroweak symmetry breaking}. By this we mean the
following: At the GUT energy scale both the Higgs mass parameters
$m_1^2$ and $m_2^2$ are positive and the Higgs potential has no
non-trivial minima. However, when running down to the EW scale due
to the radiative corrections they may change sign so that the
potential develops  a non-trivial minimum. At this minimum the
electroweak symmetry happens to be spontaneously broken. Thus,
contrary to the SM where one has to choose the negative sign of
the Higgs mass squared  "by hand", in the MSSM the effect of
spontaneous symmetry breaking is triggered by the radiative
corrections.

Indeed, one can see in Fig.\ref{16} is that $m_2^2$ (or both $m_1^2$
and $m_2^2$) decreases when going down from the GUT scale  to the
$M_Z$ scale and  can even become negative. This is the effect of
the large top (and bottom)  Yukawa couplings in the RG equations.
As a result, at some value of $Q^2$  the conditions (\ref{cond})
are satisfied, so that the non-trivial minimum appears. This
triggers spontaneous breaking of the $SU(2)$ gauge invariance. The
vacuum expectations of the Higgs fields acquire non-zero values
and provide masses to the quarks, leptons, $SU(2)$ gauge bosons,
and additional masses to their superpartners.

This way one obtains also the explanation of why the two scales are so much different.
Due to the logariphmic running of the parameters one needs  a long "running time"
to get $m_1^2$  to be negative when starting from a positive value of the order of
$M_{SUSY}$ scale $\sim 10^2 \div 10^3$ GeV.

\section{Infrared Quasi-fixed Points}
\setcounter{equation} 0

Examining the RG equations for the Yukawa couplings  one finds
that they possess the infrared fixed points. This is a very
typical behaviour for RG equations. In this section we give a
short description of the infrared quasi-fixed points
(IRQFP) \cite{Hill} in the MSSM. They play an important role in
predictions of the mass spectrum.

As in the previous section  we consider the RG flow in the  direction
from GUT to EW scale, the running parameter being $t=\log
M_{GUT}^2/Q^2$. This corresponds to the opposite sign in RG
eqs.(\ref{eq},\ref{eq2}).

\subsection{Low tan$\beta$ regime}

 Consider first the low  $\tan\beta$ regime. In this
case, the only important Yukawa coupling is the top-quark one,
all the others can be put equal to zero
and
 the RG equations can be solved analytically \cite{Ibanez}
\begin{equation}{\mbox a}_i(t) = \frac{{\mbox a}_0}{1+{\mbox a}_0b_it},
\ \ \ Y_t(t)=\frac{Y_0E_t(t)}{1+6Y_0F_t(t)},
 \label{sol}
\end{equation}
where
\begin{eqnarray*}
E_t(t)&=&\prod_i(1+b_i{\mbox a}_0t)^{\displaystyle c_{ti}/b_i} , \ \ \
 c_{ti}=(\frac{13}{15},3,\frac{16}{3}), \\
F_t(t)&=&\int^t_0 E_t(t')dt'.
\end{eqnarray*}
 In the IR regime solution (\ref{sol}) possesses a quasi-fixed
 point. Indeed taking the limit  $Y_0=Y_t (0) \to \infty$
one can drop 1 in the denominator of eq.(\ref{sol}) and obtain the
IRQFP
\begin{equation}
Y(t)\Rightarrow Y_t^{FP}=\frac{E_t(t)}{6 F_t(t)}, \label{fp} 
\end{equation}
which is independent of the initial condition \cite{Hill,10}.

\begin{figure}[t]
     \vspace{-3.5cm}
       \begin{flushleft}
       \leavevmode
       \epsfxsize=5.5cm
\epsffile{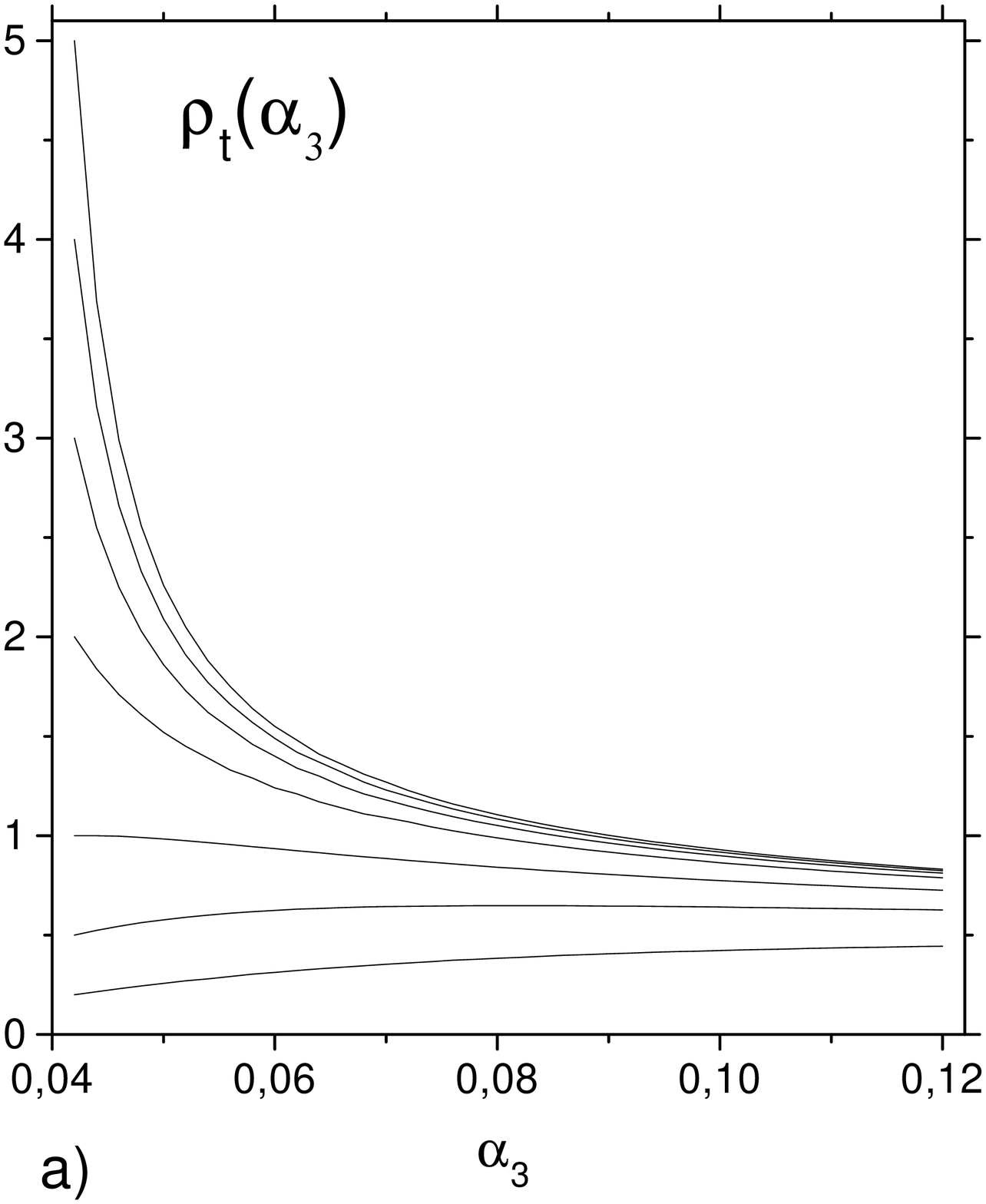}
   \end{flushleft}
     \vspace{-8.35cm}\hspace{7cm}
       \leavevmode
       \epsfxsize=5.5cm
       \epsffile{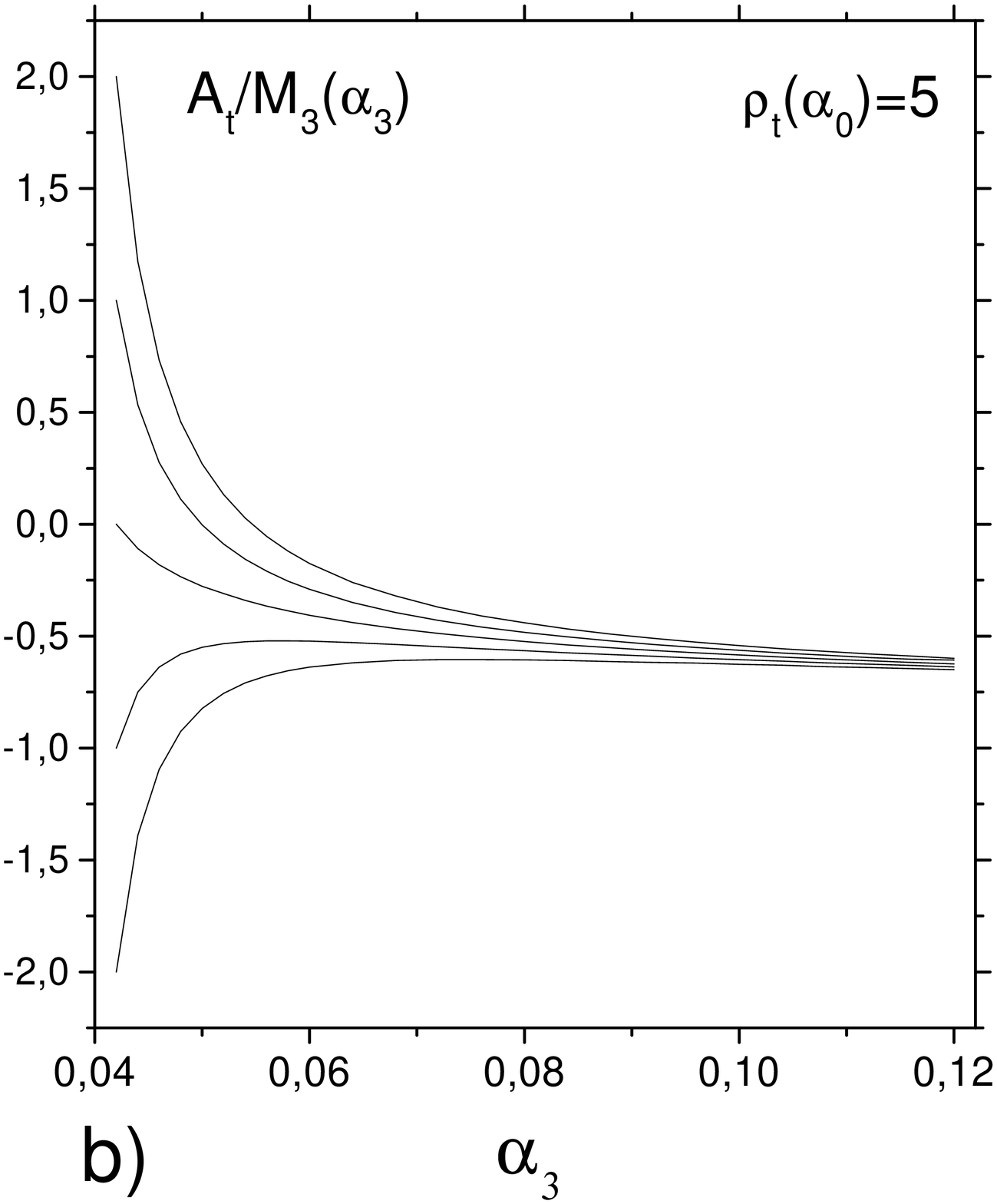}
\caption{The infrared quasi-fixed points for $\rho_t=Y_t/{\mbox a}_3
$ (a) and $\rho_A=A_t/M_3$ (b) . The x-axis shows the values of $\alpha_3$. \label{f1}}
\end{figure}

Though perturbation theory is not valid for $Y_t > 1$, it does not
prevent us from using the fixed point (\ref{fp}) since it attracts
any solution with $Y_0 > {\mbox a}_{0}$ or, numerically, for $Y_0
> 0.1/4\pi$ . Thus, for a wide range of initial values $Y_t$ is driven
to the IR quasi-fixed point given by eq. (\ref{fp}) which
numerically corresponds to $y_t(M_Z) \approx 1.125$. It is useful
to introduce the ratio $\rho_t \equiv Y_t/{\mbox a}_3$ since the
strong coupling is the leading one in the IR regime. At the fixed
point $\rho_t(M_Z) \approx 0.84$ and is approached in the IR
regime when $Q^2$ decreases. The behaviour of $\rho(t)$ is shown
in Fig.\ref{f1} \cite{YJK}.
\input epsf

To get the solutions for the soft terms it is enough to perform
 the
substitution ${\mbox a} \to \tilde{{\mbox a}}$ and $Y\to \tilde{Y}$ and
expand over $\eta $ and $\bar \eta$. Expanding the gauge coupling
in (\ref{sol}) up to $\eta$ one has (hereafter we assume
$M_{0i}=m_{1/2}$)
\begin{equation}
M_i(t)=\frac{m_{1/2}}{1+b_i{\mbox a}_0t}.
\end{equation}
Performing the same expansion for the Yukawa coupling   one finds
\cite{Kaz}
\begin{equation}
A_t(t)=\frac{A_0}{1+6Y_0F_t}-m_{1/2}\left( \frac{t}{E_t}\frac{dE_t}{dt}-
\frac{6Y_0}{1+6Y_0F_t}(tE_t-F_t) \right). \label{Aa}
\end{equation}
To get the solution for the $\Sigma$ term one has to make
expansion over $\eta$ and $\bar \eta$. This leads to
\begin{equation}
\Sigma_t(t)=\frac{\Sigma_0-A_0^2}{1+6Y_0F_t}+\frac{(A_0+m_{1/2}6Y_0(tE_t-F_t))^2}{(1+6Y_0F_t)^2}
+m_{1/2}^2\left[\frac{d}{dt}\left(\frac{t^2}{E_t}\frac{dE_t}{dt}\right)
-\frac{6Y_0}{1+6Y_0F_t}t^2\frac{dE_t}{dt}\right], \label{si}
\end{equation}

With  analytic solutions (\ref{Aa},\ref{si}) one can analyze
asymptotics and, in particular, find the infrared quasi-fixed
points which correspond to $Y_0 \to \infty$
\begin{eqnarray}
A_t^{FP}&=&-m_{1/2}\left(\frac{t}{E_t}\frac{dE_t}{dt}-\frac{tE_t-F_t}{F_t}\right),
\label{Af}\\
\Sigma^{FP}_t&=&m_{1/2}^2\left[\left(\frac{tE_t-F_t}{F_t}\right)^2+\frac{d}{dt}
\left(\frac{t^2}{E_t}\frac{dE_t}{dt}\right)-\frac{t^2}{F_t}\frac{dE_t}{dt}\right].
\label{Sf}
\end{eqnarray}
 The FP
solutions (\ref{Af},\ref{Sf}) can be directly obtained from a
fixed point for the rigid Yukawa coupling (\ref{fp}) by
Grassmannian expansion. This explains, in particular, why fixed
point solutions for the soft couplings  exist if they exist for
the rigid ones and with the same stability properties~\cite{JJP}.

The behaviour of $\rho_A=A_t/M_3$ as a function of ${\mbox a}_3$ for
a fixed ratio $Y_0/{\mbox a}_{0}=5$ is shown in Fig.\,\ref{f1}b. One
can  observe the  strong attraction to the IR stable quasi-fixed
point $\rho_A \approx -0.62$ \cite{YJK}.

One can also write down solutions for the individual masses. This
can be obtained using the Grassmannian expansion of solutions for
the corresponding superfield propagators. For the first two
generations one has
\begin{eqnarray*}
\tilde{m}^2_{Q_L}&=&m_0^2+\frac{m_{1/2}^2}{2}(\frac{16}{3}f_3+3f_2+\frac{1}{15}f_1),
\\
\tilde{m}^2_{U_R}&=&m_0^2+\frac{m_{1/2}^2}{2}(\frac{16}{3}f_3+\frac{16}{15}f_1),
\\
\tilde{m}^2_{D_R}&=&m_0^2+\frac{m_{1/2}^2}{2}(\frac{16}{3}f_3+\frac{4}{15}f_1),
\\
m^2_{H_1}&=&m_0^2+\frac{m_{1/2}^2}{2}(3f_2+\frac{3}{5}f_1),
\\
\tilde{m}^2_{L_L}&=&m_0^2+\frac{m_{1/2}^2}{2}(3f_2+\frac{3}{15}f_1),
\\
\tilde{m}^2_{E_R}&=&m_0^2+\frac{m_{1/2}^2}{2}(\frac{12}{5}f_1),
\end{eqnarray*}
where $$f_i=\frac{1}{b_i}\left(1-\frac{1}{(1+b_i{\mbox a}_0t)^2}\right).$$
 The third generation masses get the contribution from the top
 Yukawa coupling \cite{CK}
\begin{eqnarray*}
\tilde{m}^2_{b_R}&=&\tilde{m}^2_{D_R}, \\ \tilde{m}^2_{b_L}&=&
\tilde{m}^2_{Q_L}+\Delta/6, \\ \tilde{m}^2_{t_R}&=&
\tilde{m}^2_{U_R}+\Delta/3, \\ \tilde{m}^2_{t_L}&=&
\tilde{m}^2_{Q_L}+\Delta/6, \\ m^2_{H_2}&=& m^2_{H_1}+\Delta/2,
\end{eqnarray*}
where $\Delta$ is related to $\Sigma_t$ (\ref{si}) by $$ \Delta =
\Sigma_t-\Sigma_0-m_{1/2}^2\frac{d}{dt}\left(\frac{t^2}{E_t}\frac{dE_t}{dt}\right).$$

\begin{figure}[ht] \vspace{-0.8cm}
 \begin{flushleft}
    \leavevmode
    \epsfxsize=5.5cm
    \epsffile{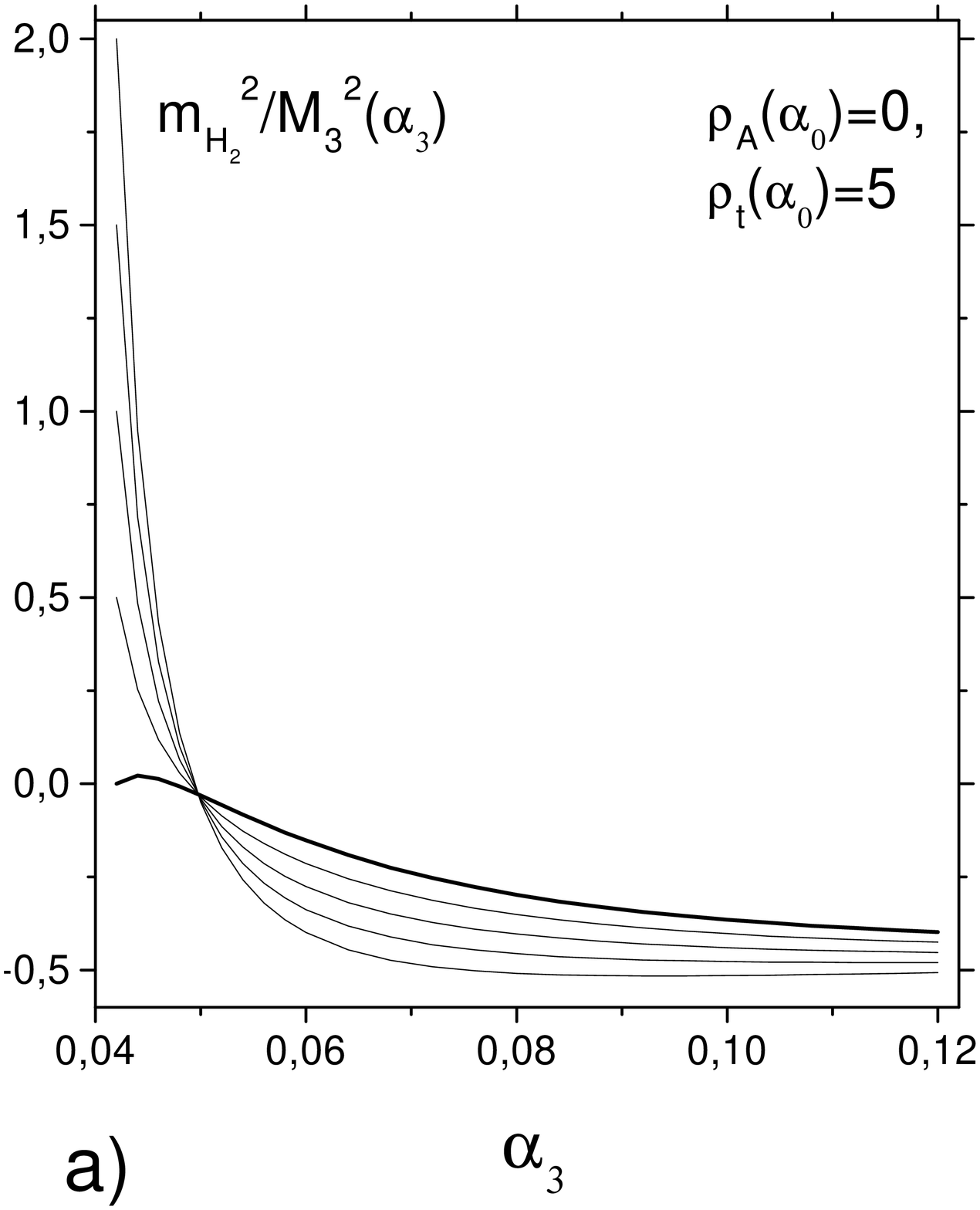}
  \end{flushleft}
\vspace{-8.35cm}
 \hspace{7cm}
    \leavevmode
    \epsfxsize=5.5cm
    \epsffile{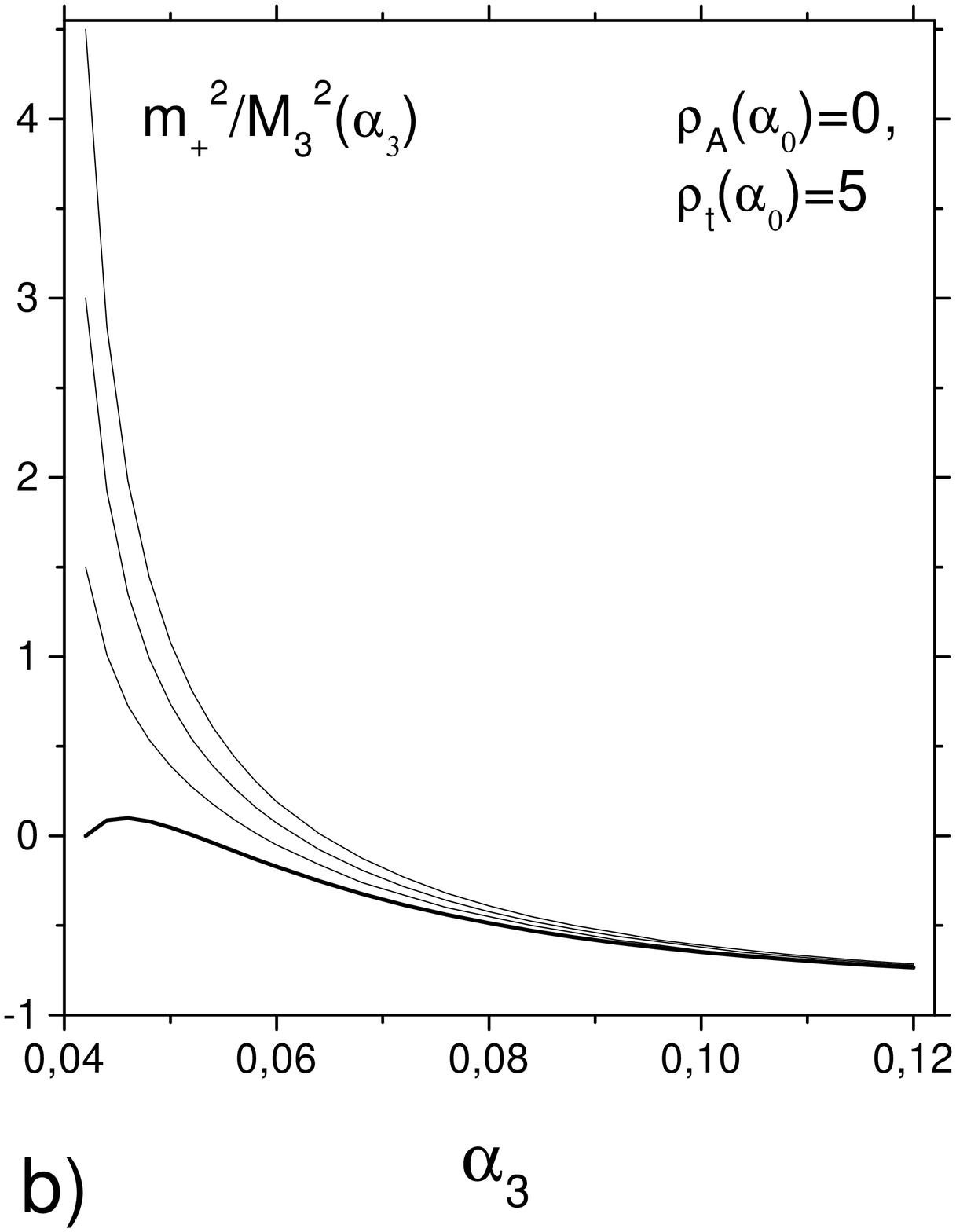}
\caption{The infrared quasi-fixed points for $m_{H_2}^2/M_3^2$ (a)
and  $(m_{H_1}^{2}+2 m_{H_2}^2)/M_3^2$ (b) \label{f2}}
  \end{figure}

There is no obvious infrared attractive fixed point for
$m_{H_1}^2$.  However, one can take the  linear combination
$m_+^2=m_{H_1}^2+2 m_{H_2}^2$  which together with $m_{H_2}^2$
shows the IR fixed point behaviour in the limit  $Y_0\to \infty$.
\begin{equation} \frac{m_+^{2\ FP}}{M_3^2}
\approx -0.73\, ,\ \ \ \ \frac{m_{H_2}^{2\ FP}}{M_3^2} \approx -0.12
\left(\frac{1}{2}\frac{m_0^2}{m_{1/2}^2}+3.4 \right) \,. \label{h1}
\end{equation}
In eq.(\ref{h1}) one has only  weak dependence on the ratio
$m_0^2/m_{1/2}^2$.  One can find the IR quasi-fixed point
$m_{H_2}^2/M_3^2 \approx -0.40$ which corresponds to
$m_0^2/m_{1/2}^2=0$.  As for the combination $m_+^2$, the
dependence on initial conditions disappears completely, as it
follows from (\ref{h1}). The situation is illustrated in
Fig.\ref{f2} \cite{YJK}.
\input epsf

\input epsf
    \begin{figure}[ht]
   \vspace{-1cm}
    \begin{flushleft}
     \leavevmode
    \epsfxsize=5.5cm
    \epsffile{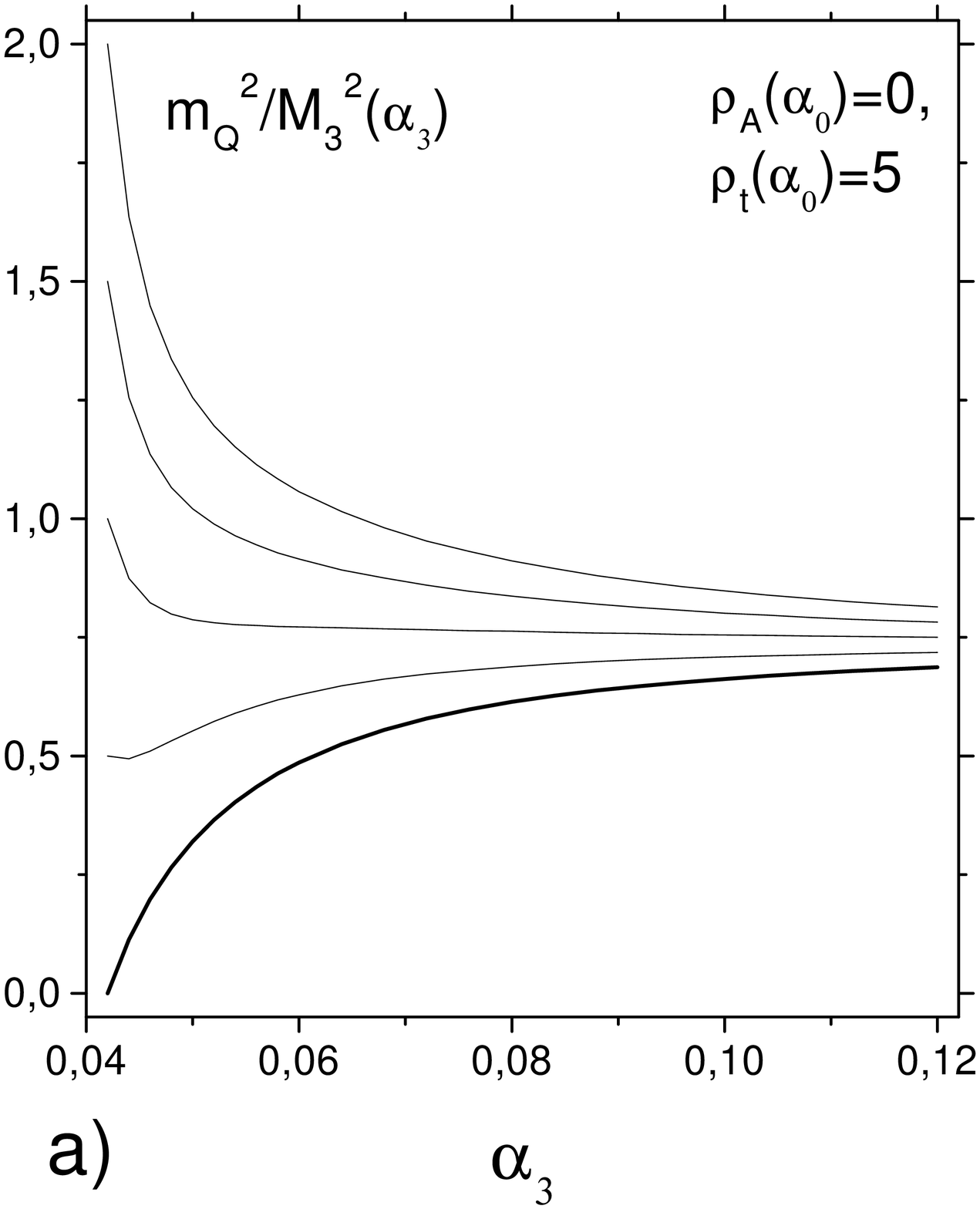}
  \end{flushleft}
\vspace{-8.35cm} \hspace{7cm}
    \leavevmode
    \epsfxsize=5.5cm
    \epsffile{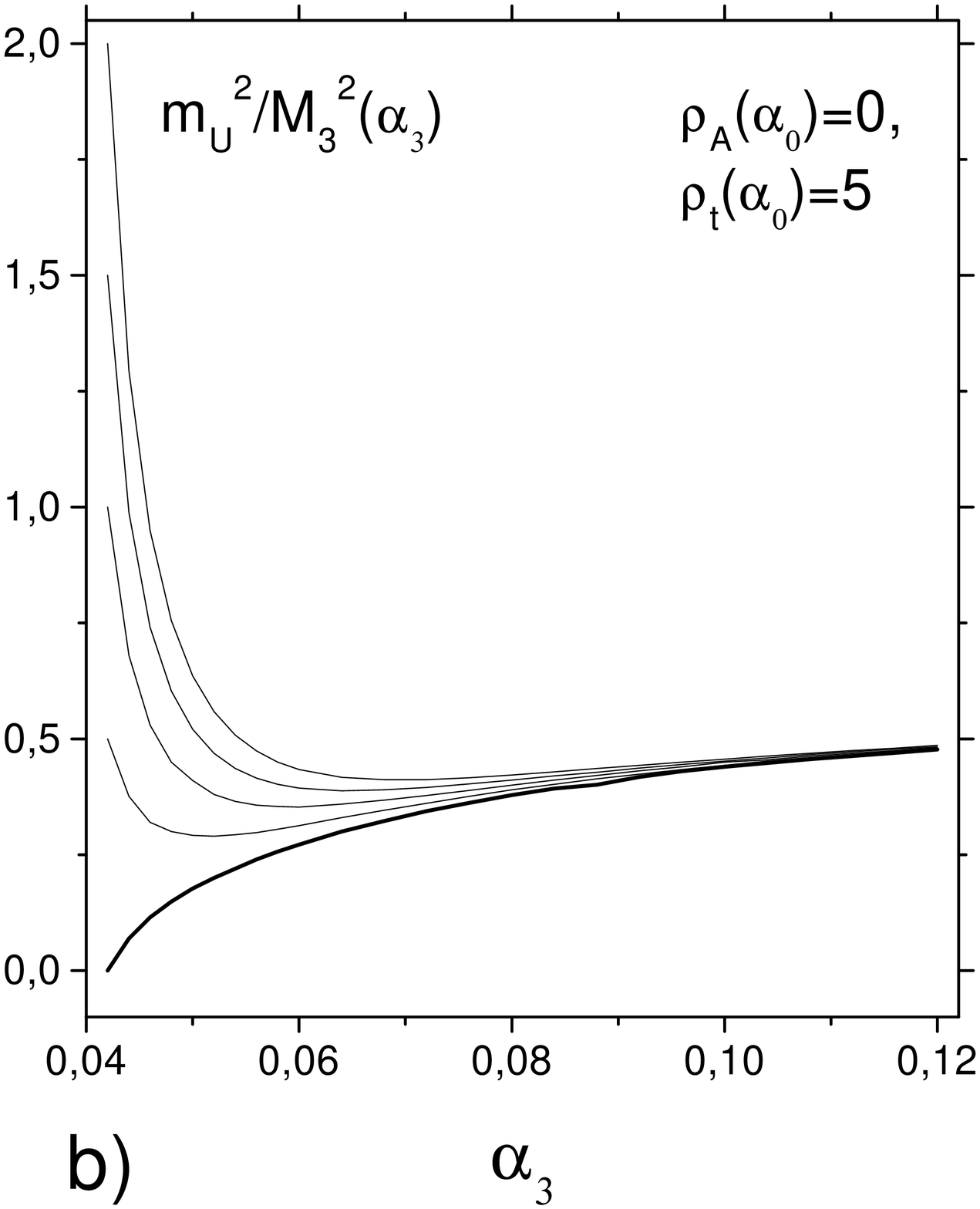}
\caption{The infrared quasi-fixed points for
$\tilde{m}_{Q}^2/M_3^2$ (a) and  $\tilde{m}_U^{2}/M_3^2$ (b)
\label{f3}}
  \end{figure}
  Consider  now the squark masses. In the limit $Y_t \to \infty$
these solutions are driven to the IRQFP's
\begin{equation}
\frac{\tilde{m}_U^{2\ FP}}{M_3^2} \approx 0.48 , \ \ \ \frac{\tilde{m}_Q^{2\
FP}}{M_3^2}  \approx 0.12 \left( \frac{1}{2} \frac{m_0^2}{m_{1/2}^2} + 5.8
\right) \label{mqfp}
\end{equation}
As it follows from eq.(\ref{mqfp}), the solution for
$\tilde{m}_{U}^2/M_3^2$ becomes independent of the initial
conditions $m_0/m_{1/2}$  and $A_0/m_{1/2}$, when the top-quark
Yukawa coupling is initially large enough. As a result, the
solutions of RGE's are driven to the fixed point (\ref{mqfp}) for
a wide range of $m_0^2/m_{1/2}^2$ (Fig.\ref{f3}b). As for $m_Q^2$,
the dependence on initial conditions does not completely
disappear; however, it is rather weak like in the case of
$m_{H_2}^2$ and approaches the value $\tilde{m}_{Q}^2/M_3^2
\approx 0.69$ (Fig.\ref{f3}a) \cite{YJK}.

The bilinear SUSY breaking parameter B does not exhibit a
fixed-point behaviour in the limit $Y_0 \gg {\mbox a}_{0}$. To see
this, consider the solution in the aforementioned limit. One has
\begin{equation} \frac{B}{M_3}
  \approx
0.35  \left(\frac{B_0}{m_{1/2}} - \frac{1}{2} \frac{A_0}{m_{1/2}}
-0.8 \right). \label{B}
\end{equation}
It is clear that neither $B_0/m_{1/2}$ nor $A_0/m_{1/2}$ may  be
neglected. As a consequence, no fixed point behaviour for the
ratio $B/M_3$  is observed.

We have considered the one-loop  RGE's. It is interesting to see,
however, how our results are modified when two-loop RGE's are
used. For comparison we present the two-loop IRQFP values
\cite{AA} together with our one-loop results in Table 1.

\begin{table}[ht]
\begin{center}
\begin{tabular}{|c|c|c|c|c|c|}
\hline
 & $A_t/M_3$  & $\tilde{m}_U^2/M_3^2$ & $m_+^2/M_3^2$ & $m_{H_2}^2/M_3^2$ &
$\tilde{m}_Q^2/M_3^2$ \\ \hline one-loop level & -0.62 & 0.48 &
-0.73 & -0.40 & 0.69 \\ \hline two-loop level & -0.59 & 0.48 &
-0.72 & -0.46 & 0.75 \\ \hline
\end{tabular} \end{center}
\caption{ Comparison of one- and two-loop IRQFP's}\label{taa}
\end{table}

As one can see from this table, the difference between the
one-loop and two-loop  results is negligible for $A_t/M_3$,
$\tilde{m}_U^2/M_3^2$ and $m_+^2/M_3^2$. As for $m_{H_2}^2/M_3^2$
and $\tilde{m}_Q^2/M_3^2$, the two-loop corrections to the fixed
points are about two times as small as deviations from them. As it
was mentioned above, such  corrections  have a negligible impact
on our main results.

\subsection{Large tan$\beta$ regime}

We now give a short description of the infrared behaviour of the
RGE's in the MSSM for the large $\tan\beta$ regime. While with a
single Yukawa coupling the analytical solution to the one-loop RG
equations has been known for long, for increasing number of Yukawa
couplings it has been obtained quite recently \cite{AM} in the
form that allows iterative representation.

 One can write down the one-loop RG equations (\ref{eq}) as
\begin{eqnarray}
\frac{d{\mbox a}_i}{dt} &=& b_i{\mbox a}_i^2, \label{al}\\ \frac{dY_k}{dt} &=&
-Y_k(\sum_{i}c_{ki}{\mbox a}_i - \sum_{l}a_{kl}Y_l), \label{y}
\end{eqnarray}
where
\begin{eqnarray*}
b_i&=&\{33/5,1,-3 \}, \\ c_{ti}&=& \{13/15,3,16/3 \}, \ \
c_{bi}=\{7/15,3,1 6/3 \},\ \  c_{\tau i}=\{9/5,3,0 \}, \\
a_{tl}&=&\{6,1,0 \},\ \ a_{bl}=\{1,6,1 \},\ \ a_{\tau l}=\{0,3,4
\}.
\end{eqnarray*}
Then the general solution to eqs.(\ref{al},\ref{y}) can be written in the form
\cite{AM}
\begin{eqnarray}
{\mbox a}_i&=& \frac{{\mbox a}_0}{1+b_i{\mbox a}_0t},
 \label{sola}\\
Y_k & =&  \frac{Y_k^0u_k}{1+a_{kk}Y_k^0\int_0^t u_k},
 \label{soly}
\end{eqnarray}
where the functions $ u_k$ obey the integral system of equations
\begin{equation}
u_t=\frac{E_t}{(1+6Y_b^0\int_0^t u_b)^{1/6}}, \ \
u_b=\frac{E_b}{(1+6Y_t^0\int_0^t u_t)^{1/6} (1+4Y_\tau^0\int_0^t
u_\tau)^{1/4}}, \ \  u_\tau=\frac{E_\tau}{(1+6Y_b^0\int_0^t
u_b)^{1/2}}, \label{u}
\end{equation}
and the functions $E_k$ are given by
\begin{equation}
E_k= \prod_{i=1}^3(1+b_i{\mbox a}_0t)^{c_{ki}/b_i}. \label{e}
\end{equation}

Let us stress that eqs.(\ref{sola},\ref{soly}) give the exact solution to
eqs.(\ref{al},\ref{y}), while the $u_k$'s in eqs.(\ref{u}), although solved
formally in terms of the $E_k$'s and $Y_k^0$'s as continued integrated
fractions, should in practice be solved iteratively.

Let us now perform the substitution (\ref{ga},\ref{ya}) in
(\ref{sola}-\ref{u}) and expand over $\eta$ and $\bar \eta$. Then
the linear term in $\eta$ will give us the solution for $M_i$ and
$A_k$ and the $\eta \bar \eta$ terms the ones for $\Sigma_k$.  The
resulting exact solutions look similar to those for the rigid
couplings (\ref{sola}--\ref{u}) \cite{KM}
\begin{eqnarray}
M_i&=& \frac{M_i^0}{1+b_i{\mbox a}_i^0t}, \label{m} \\
 A_k &=& -e_k + \frac{A_k^0/Y_k^0 +a_{kk}\int
u_ke_k}{1/Y_k^0 +a_{kk}\int u_k},\label{Ak}  \\ \Sigma_k &=&
\xi_k+A_k^2+2e_kA_k -\frac{(A_k^0)^2/Y_k^0
-\Sigma_k^0/Y_k^0+a_{kk}\int u_k\xi_k}{1/Y_k^0+a_{kk}\int u_k},
\label{S}
\end{eqnarray}
where the new functions $e_k$ and $\xi_k$ have been introduced
which obey the iteration equations
\begin{eqnarray}
 e_t &=& \frac{1}{E_t}\frac{d\tilde{E}_t}{d\eta}+ \frac{A_b^0\int u_b-\int
u_be_b}{ 1/Y_b^0+6\int u_b } , \nonumber
\\
 e_b &=& \frac{1}{E_b}\frac{d\tilde{E}_b}{d\eta} +\frac{A_t^0\int
u_t-\int u_te_t}{ 1/Y_t^0+6\int u_t} + \frac{A_\tau^0\int
u_\tau-\int u_\tau e_\tau}{1/Y_\tau^0 +4\int u_\tau} , \nonumber
\\
 e_\tau &=& \frac{1}{E_\tau}\frac{d\tilde{E}_\tau}{d\eta}+3\frac{A_b^0\int u_b
 -\int u_be_b}{1/Y_b^0+6\int u_b}, \nonumber
\\
&&   \nonumber \\
 \xi_t &=&
\frac{1}{E_t}\frac{d^2\tilde{E}_t}{d\eta d\bar\eta}
 +2\frac{1}{E_t}\frac{d\tilde{E}_t}{d\eta} \frac{A_b^0\int u_b-\int
u_be_b}{1/Y_b^0+6\int u_b} + 7\left(\frac{A_b^0\int u_b-\int
u_be_b}{ 1/Y_b^0 +6\int u_b} \right)^2  \nonumber
\\
 &&-\left((\Sigma_b^0+(A_b^0)^2)\int u_b
-2A_b^0\int u_be_b +\int u_b\xi_b\right)/\left( \frac{1}{Y_b^0}
+6\int u_b \right), \nonumber \\
 \xi_b &=&
\frac{1}{E_b}\frac{d^2\tilde{E}_b}{d\eta d\bar\eta}
+2\frac{1}{E_b}\frac{d\tilde{E}_b}{d\eta}\left[ \frac{A_t^0\int
u_t-\int u_te_t}{1/Y_t^0 +6\int u_t} +\frac{A_\tau^0\int
u_\tau-\int u_\tau e_\tau}{ 1/Y_\tau^0 +4\int u_\tau}
\right]\nonumber \\ && +7\left(\frac{A_t^0\int u_t-\int u_te_t}{
1/Y_t^0 +6\int u_t} \right)^2  +5\left(\frac{A_\tau^0\int
u_\tau-\int u_\tau e_\tau}{1/Y_\tau^0 +4\int u_\tau} \right)^2
\nonumber \\ && +2\left(\frac{A_t^0\int u_t-\int u_te_t}{1/Y_t^0
+6\int u_t}\right) \left(\frac{A_\tau^0\int u_\tau-\int u_\tau
e_\tau}{ 1/Y_\tau^0 +4\int u_\tau}\right)\nonumber \\
 &&-\left((\Sigma_t^0+(A_t^0)^2)\int u_t -2A_t^0\int u_te_t
+\int u_t\xi_t\right)/\left( \frac{1}{Y_t^0}+6\int u_t \right)
\nonumber \\
 && -\left((\Sigma_\tau^0+(A_\tau^0)^2)\int u_\tau
-2A_\tau^0\int u_\tau e_\tau +\int u_\tau\xi_\tau\right)/\left(
\frac{1}{Y_\tau^0} +4\int u_\tau \right) , \nonumber \\
 \xi_\tau &=&
 \frac{1}{E_\tau}\frac{d^2\tilde{E}_\tau}{d\eta d\bar\eta}
 +6\frac{1}{E_\tau}\frac{d\tilde{E}_\tau}{d\eta} \frac{A_b^0\int u_b-\int
u_be_b}{1/Y_b^0+6\int u_b} +27\left(\frac{A_b^0\int u_b-\int
u_be_b}{1/Y_b^0+6\int u_b} \right)^2 \nonumber\\
 &&-3 \left((\Sigma_b^0+(A_b^0)^2)\int u_b-2A_b^0\int u_be_b
 +\int u_b\xi_b\right)/\left( \frac{1}{Y_b^0}+6\int u_b \right).
\label{ex}
\end{eqnarray}
Here the variations of $\tilde{E}_k$ should be taken at $\eta = \bar
\eta=0$. When solving eqs.(\ref{u}) and (\ref{ex}) in the $n$-th iteration
one has to substitute in the r.h.s. the $(n-1)$-th iterative solution for
all the corresponding functions.

The solutions for the individual soft masses are linearly
expressed through $\Sigma$'s \cite{CK}
\begin{eqnarray*}
\tilde{m}^2_{Q_3}&=&m_0^2+m_{1/2}^2\frac{128f_3+87f_2-11f_1}{122}
+\frac{17(\Sigma_t-\Sigma_t^0)+20(\Sigma_b-\Sigma_b^0)
-5(\Sigma_\tau-\Sigma_t^0)}{122},\nonumber \\
\tilde{m}^2_{U_3}&=&m_0^2+m_{1/2}^2\frac{144f_3-108f_2+144/5f_1}{122}
+\frac{42(\Sigma_t-\Sigma_t^0)-8(\Sigma_b-\Sigma_b^0)
+2(\Sigma_\tau-\Sigma_t^0)}{122},\nonumber\\
\tilde{m}^2_{D_3}&=&m_0^2+m_{1/2}^2\frac{112f_3-84f_2+112/5f_1}{122}
+\frac{-8(\Sigma_t-\Sigma_t^0)+48(\Sigma_b-\Sigma_b^0)
-12(\Sigma_\tau-\Sigma_t^0)}{122},\nonumber\\
{m}^2_{H_1}&=&m_0^2+m_{1/2}^2\frac{-240f_3-3f_2-57/5f_1}{122}
+\frac{-9(\Sigma_t-\Sigma_t^0)+54(\Sigma_b-\Sigma_b^0)
+17(\Sigma_\tau-\Sigma_t^0)}{122},\nonumber\\
{m}^2_{H_2}&=&m_0^2+m_{1/2}^2\frac{-272f_3+21f_2-89/5f_1}{122}
+\frac{63(\Sigma_t-\Sigma_t^0)-12(\Sigma_b-\Sigma_b^0)
+3(\Sigma_\tau-\Sigma_t^0)}{122},\nonumber\\
\tilde{m}^2_{L_3}&=&m_0^2+m_{1/2}^2\frac{80f_3+123f_2-103/5f_1}{122}
+\frac{3(\Sigma_t-\Sigma_t^0)-18(\Sigma_b-\Sigma_b^0)
+35(\Sigma_\tau-\Sigma_t^0)}{122},\nonumber\\
\tilde{m}^2_{E_3}&=&m_0^2+m_{1/2}^2\frac{160f_3-120f_2+32f_1}{122}
+\frac{6(\Sigma_t-\Sigma_t^0)-36(\Sigma_b-\Sigma_b^0)
+70(\Sigma_\tau-\Sigma_t^0)}{122}. \label{masses}
\end{eqnarray*}

The solutions (\ref{sola}--\ref{u}, \ref{m}--\ref{ex}) have a nice
property since they contain explicit dependence on initial
conditions and one can trace this dependence in the final results.
This is of special importance for the non-universal case since one
can see which of the parameters is essential and which is  washed
out during the evolution. In particular the solution for the
Yukawa couplings exhibit the fixed point behaviour when the
initial values are large enough. More precisely, in the regime
$Y_t^0, Y_b^0, Y_\tau^0 \to \infty$ with fixed finite ratios
$Y_t^0/Y_b^0= r_1, Y_b^0/Y_\tau^0= r_2$, it is legitimate to drop
$1$ in the denominators of eqs.(\ref{soly}, \ref{u})  in which
case the exact Yukawa solutions go to the IRQFP defined by
\cite{KM}
\begin{equation}
Y_k^{FP} =  \frac{u_k^{FP}}{a_{kk}\int u_k^{FP}} \label{fp2}
\end{equation}
with
\begin{equation}
u_t^{FP}=\frac{E_t}{(\int u_b^{FP})^{1/6}}, \ \
 u_b^{FP}=\frac{E_b}{(\int u_t^{FP})^{1/6} (\int u_\tau^{FP})^{1/4}} , \ \
 u_\tau^{FP}=\frac{E_\tau}{(\int u_b^{FP})^{1/2}}
\label{ufp}
\end{equation}
extending the IRQFP (\ref{fp}) to three Yukawa couplings. What is worth
stressing here is that both the dependence on the initial condition for each
Yukawa as well as the effect of Yukawa non-unification, ($r_1, r_2$),  have
completely dropped out of the runnings.

 This in turn leads to the IRQFPs for the soft
terms. Disappearance of $Y_k^0$ in the FP solution naturally leads
to the disappearance of $A_k^0$ and $\Sigma_k^0$ in the soft term
fixed points.

Below we present the result of numerical analysis. We begin with
Yukawa couplings and assume the equality  of the Yukawa couplings
of the third generation at the GUT scale:
$Y_t(M_{GUT})=Y_b(M_{GUT})=Y_{\tau}(M_{GUT})$.

In Figs.\,\ref{f1}a,b,c  the numerical solutions of the RGE's are
shown for a wide range of initial values of
$\rho_t(M_{GUT})=\rho_b(M_{GUT})=\rho_{\tau}(M_{GUT})$ from the
interval $<0.2, 5>$, where $\rho_i=Y_i/{\mbox a}_3$. One can
clearly see the IRQFP type  behaviour when the parameter $\rho_i$
at the GUT scale is big enough \cite{JK}.

We have found the following values of the Yukawa couplings $y_i$
at the $M_Z$ scale $$y_t \in <0.787, 1.048>, \ \ y_b \in <0.758,
0.98>,\ \ \ y_{tau} \in <0.375, 0.619>.$$
 Comparing $y_t$ and $y_b$  one can see that the ratio
belongs to a very narrow interval $y_t/y_b \in <1.039, 1.069>$.

Now we proceed with the discussion of  RGE's for  trilinear scalar
couplings, $A_i, i=(t, b, \tau)$.  The results are shown in
Figs.\,\ref{f1}d,e for the following quantities
$\rho_{A_i}=A_i/M_3, i=(t, b)$ for different initial values at the
GUT scale and for $\rho_i(M_{GUT})=5$.  One can  see the strong
attraction to the fixed points \cite{JK}.

\begin{figure}[t]
     \vspace{-11.5cm}
       \begin{flushleft}
       \leavevmode
       \epsfxsize=17cm
       \epsffile{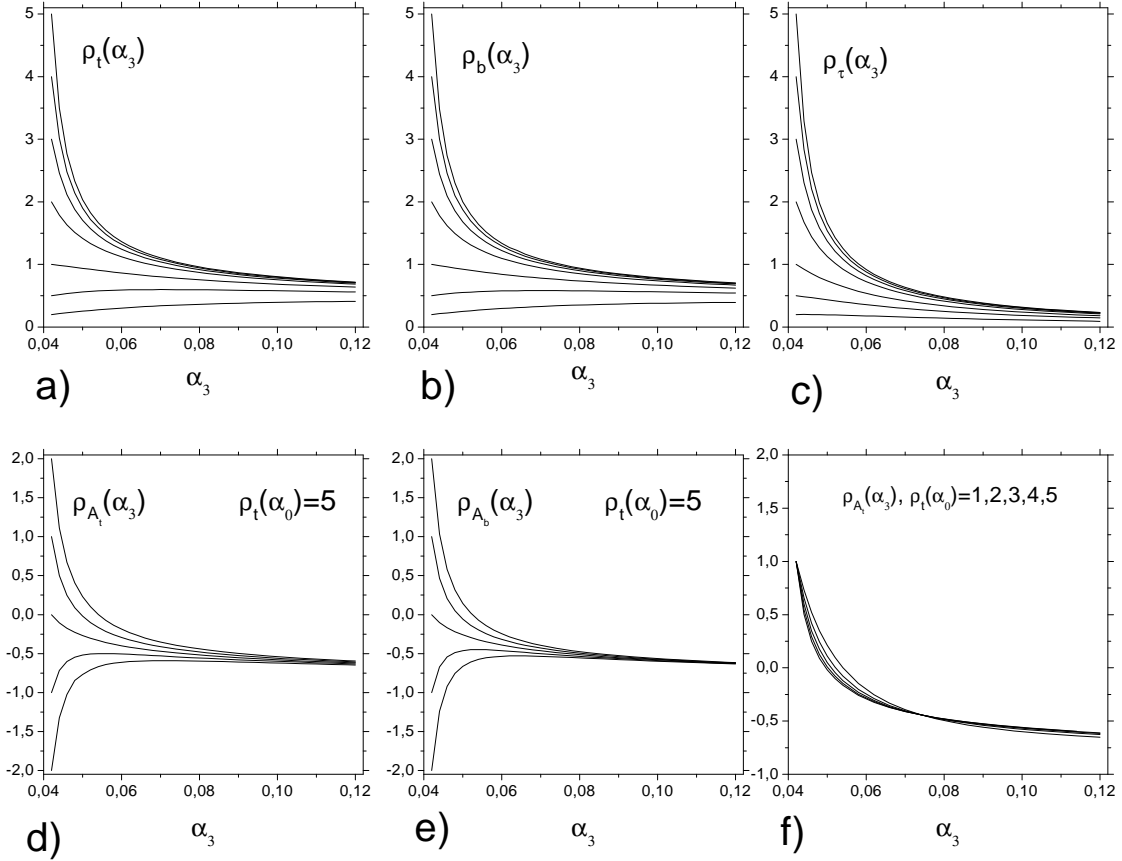}
   \end{flushleft}
\caption{The infrared quasi fixed points for $\rho_i=Y_i/{\mbox
a}_3$ $i=t,b,\tau$ (a,b,c), $\rho_{A_i}=A_i/ M_3$ $i=t,b$ (d,e)
and $\rho_{A_t}$ with $\rho_{A_t}({\mbox a}_0)=1$ for different
initial values of $\rho_t({\mbox a}_0)$ (f). \label{f1e}}
\end{figure}

The question of stability of these IRQFPs becomes important for
further consideration. Analyzing their stability under the
change of the initial conditions for $\rho_i(M_{GUT})$ one
finds remarkable stability, which allows to use them as  fixed
parameters at the $M_Z$ scale. In Fig.\,\ref{f1e}f  a particular
example of stability of IRQFP for $A_t$ is shown. As a result one
has the following IRQFP values for the parameters $\rho_{A_i}$:
 $$\rho_{A_t}\approx -0.619, \ \ \rho_{A_b}\approx
-0.658, \ \ \rho_{A_{\tau}}\approx 0.090.$$

The last step in the investigation of the RGE's is the calculation
of the soft mass parameters.
\input epsf
\input epsf
   \begin{figure}[t]
     \vspace{-11.5cm}
       \begin{flushleft}
       \leavevmode
       \epsfxsize=17cm
       \epsffile{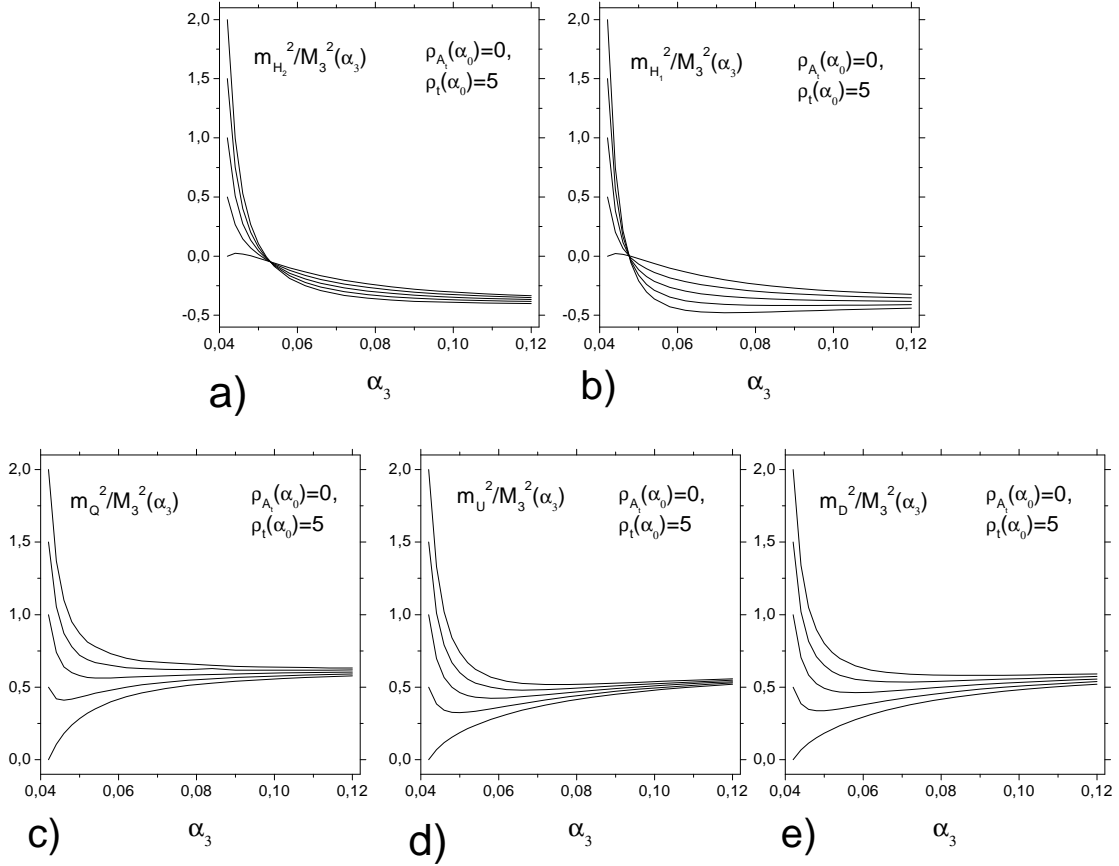}
   \end{flushleft}
\caption{The infrared quasi fixed points for mass parameters.
\label{f2e}}
\end{figure}

\noindent As one can see from Figs. \ref{f2e}a,b there exist  IRQFP's
\cite{JK}
 $$m_{H_1}^2/M_3^2 \approx-0.306,\  \ \ m_{H_2}^2/M_3^2 \approx -0.339.$$
The numbers correspond to the initial condition
$m_0^2/m_{1/2}^2=0$. Later the  initial values for the
ratio $m_0^2/m_{1/2}^2$ belonging to the following interval
$m_0^2/m_{1/2}^2 \in <0, 2>$ are considered.

In Figs.\,\ref{f2e}c,d,e the infrared behaviour of the soft SUSY
breaking squark masses is shown. One can immediately see that all
masses have IRQFPs which are used in the next section to find the
mass spectrum. For further analysis only the squark masses are
important. As for sleptons they also have an attractive infrared
behaviour but it does not influence the mass spectrum of the Higgs
bosons in which we are interested in below and we do not show them
explicitly.

Numerical values of the ratios are the following \cite{JK}:
$$m_Q^2/M_3^2 \approx 0.58, \ \ m_U^2/M_3^2 \approx 0.52, \ \
m_D^2/M_3^2 \approx 0.53,$$ obtained for $m_0^2/m_{1/2}^2=0$. One
again finds a very week dependence on initial values of the Yukawa
couplings.

The behaviour of the bilinear SUSY breaking parameter $B$
 is the same as in low $\tan\beta$ case. The ratio
$B/M_3$ does not exhibit the infrared quasi fixed point behaviour.

Thus, one can see that solutions of RGE's for all MSSM SUSY
breaking parameters (the only exception is the parameter $B$) are
driven to the infrared attractive fixed points if  the Yukawa
couplings at the GUT scale are large enough.

Our analysis is constrained by the one-loop RG equations.  The
difference between one-loop and two-loop IRQFPs is similar to the
low $\tan\beta$ case (see Table \ref{taa}) and is less than 10 per
cent. At the same time the deviations from the IRQFPs obtained by
one-loop RGEs are also of the same order which defines the
accuracy of our predictions. The only place where it really
matters is the prediction of the lightest Higgs boson mass where all the
proper corrections are taken into account.

\section{Higgs Boson Mass Prediction in the SM and MSSM}
\setcounter{equation} 0

\subsection{The Higgs boson mass in the SM}

The last unobserved particle from the Standard Model is the Higgs
boson. Its discovery would allow one to complete the SM paradigm
and confirm the mechanism of spontaneous symmetry breaking. On the
contrary, the absence of the Higgs boson would awake doubts about
the whole picture and would require new concepts.

\begin{figure}[ht]
\begin{center}\vspace{-1cm}
\epsfig{figure=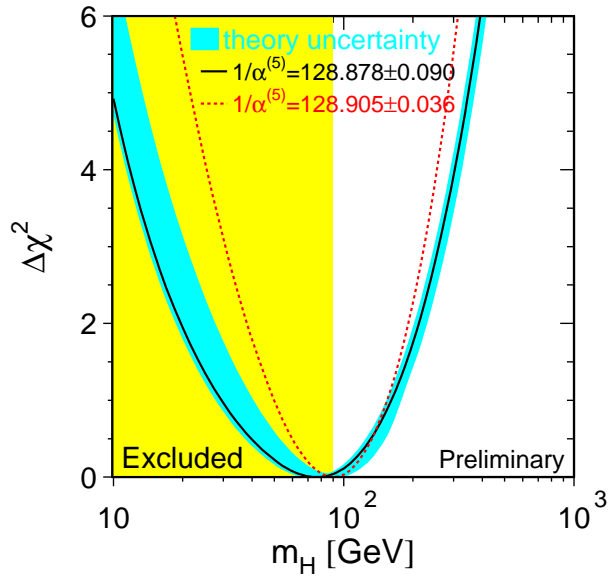, height=9cm}
\end{center}\vspace{-1cm}
\caption{The $\chi^2$ distribution as function of the Higgs mass
from the SM fit to the electroweak precision observables and the
top mass. The shaded area is excluded by the direct searches.}
\label{fig:1}
\end{figure}

Experimental limits on the Higgs boson mass come from a direct
search at LEP II and Tevatron and from indirect fits of
 electroweak precision data, first of all from the radiative
corrections to the W and top quark masses. A combined fit of
modern experimental data gives~\cite{EWWG}
\begin{equation}
 m_h=78^{+86}_{-47}\; {\rm GeV},
\end{equation}
 which at the 95\% confidence level leads to
the upper bound of 260 GeV (see Fig.\ref{fig:1}). At the same time, recent
direct searches at LEP II for the c.m. energy of 189 GeV give the lower
limit of almost 95 GeV\cite{EWWG}\footnote{The last run of LEP II at 200 GeV
c.m. energy has increased this bound up to 103 GeV\cite{LEPNEW}} . From
theoretical point of view low Higgs mass could be a hint for  physics
beyond the SM, in particular for the  supersymmetric extension of the SM.

Within the Standard Model the value of the Higgs mass $m_h$ is not
predicted. However, one can get the bounds on the Higgs mass
\cite{bound1, bound2}. They follow from the behaviour of the
quartic coupling which obeys the RG equation (\ref{lambda}).

 Since the quartic coupling grows with rising energy
indefinitely, an upper bound on $m_h$ follows from the requirement
that the theory be valid up to the scale $M_{Planck}$ or up to a
given cut-off scale $\Lambda$ below $M_{Planck}$ \cite{bound1}.
The scale $\Lambda$ could be identified with the scale at which a
Landau pole develops.  The upper bound on $m_h$ depends mildly on
the top-quark mass through the impact of the top-quark Yukawa
coupling on the running of the quartic coupling $\lambda$.

On the other hand, the requirement of vacuum stability in the SM
(positivity of $\lambda$) imposes a lower bound on the Higgs boson
mass, which crucially
 depends on the top-quark mass as well as on the cut-off $\Lambda$
  \cite{bound1,bound2}.
Again, the dependence of this lower bound on $m_t$ is due to the
effect of the top-quark Yukawa coupling on the quartic coupling in
eq.(\ref{lambda}), which drives $\lambda$ to negative values at
large scales, thus destabilizing the standard electroweak vacuum.

From the point of view of LEP and Tevatron physics, the upper
bound on the SM Higgs boson mass does not pose any relevant
restriction. The lower bound on $m_h$, instead, is particularly
important in view of search for the Higgs boson at LEPII and
Tevatron. For $m_t\sim  174$ GeV and $\alpha_s(M_Z)=0.118$ the
results at $\Lambda=10^{19}$ GeV or at $\Lambda=1$ TeV can be
given by the approximate formulae \cite{bound2}
\begin{eqnarray}
m_h&>&135
+2.1[m_t-174]-4.5\left[\frac{\alpha_s(M_Z)-0.118}{0.006}\right], \
\ \ \ \Lambda=10^{19} \  GeV, \label{19G}\\ m_h&>&72
+0.9[m_t-174]-1.0\left[\frac{\alpha_s(M_Z)-0.118}{0.006}\right], \
\ \ \ \Lambda=1 \ TeV, \label{1T}
\end{eqnarray}
where the masses are in units of  GeV.

\begin{figure}[htb]
\begin{center}\vspace{-0.3cm}
\epsfig{file=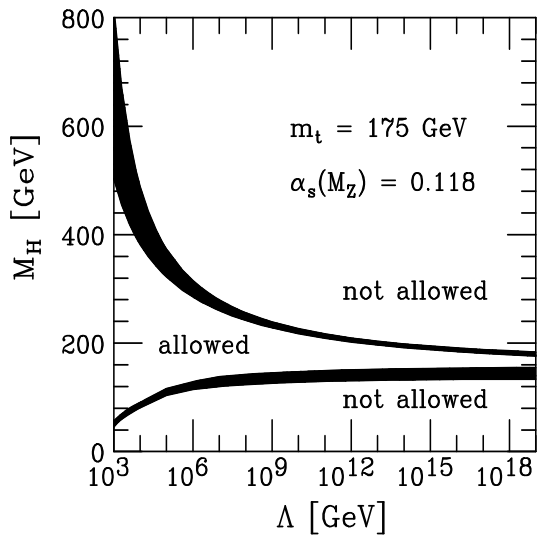,height=7cm,width=7cm,angle=90}
\end{center}
\caption{ Strong interaction and stability bounds on the SM Higgs
  boson mass. $\Lambda$ denotes the energy scale up to which the SM is valid.}
\label{mariano}
\end{figure}

Fig.\ref{mariano}~\cite{HR} shows the perturbativity and stability
bounds on the Higgs boson mass of the SM for different values of
the cut-off $\Lambda$ at which new physics is expected.
%
%
We see from Fig.\ref{mariano} and eqs.(\ref{19G},\ref{1T}) that
indeed for $m_t\sim 174$ GeV the discovery of a Higgs particle at
LEPII would imply that the Standard Model breaks down at a scale
$\Lambda$ well below $M_{GUT}$ or $M_{Planck}$, smaller for
lighter Higgs.  Actually, if the SM is valid up to $\Lambda \sim
M_{GUT}$ or $M_{Planck}$, for $m_t\sim 174$ GeV only a small range
of values is allowed: $134<m_h<\sim 200$ GeV.  For $m_t$ = 174 GeV
and $m_h < 100$~GeV [i.e. in the LEPII range] new physics should
appear below the scale $\Lambda \sim$ a few to 100~TeV. The
dependence on the top-quark mass however is noticeable. A lower
value, $m_t \simeq$ 170 GeV, would relax the previous requirement
to $\Lambda \sim 10^3 $ TeV, while a heavier value $ m_t \simeq$
180 GeV would demand new physics at an energy scale as low as
10~TeV.

\subsection{The Higgs boson mass in the MSSM}

It has been already mentioned that in the MSSM the mass of the
lightest Higgs boson is predicted to be less than the $Z$-boson mass.
This is, however, the tree level result and the masses acquire the
radiative corrections.

With account of the radiative corrections the effective Higgs bosons potential is
\begin{equation}
V_{Higgs}^{eff} = V_{tree} + \Delta V,
\end{equation}
where $V_{tree}$ is given by eq.(\ref{Higpot}) and in one loop
order
\begin{equation}
 \Delta V_{1 loop}= \sum_k \frac{1}{64\pi^2}(-1)^{J_k}(2J_k+1)c_km_k^4
 \left(\log \frac{m_k^2}{Q^2}- \frac{3}{2}\right).
\end{equation}
Here the sum is taken  over all the particles in the loop, $J_k$
is the spin and $m_k$ is the field dependent mass of a particle at
the scale $Q$.
 These radiative corrections vanish when
supersymmetry is not broken and are positive in softly broken
case.
The leading contribution comes from (s)top loops
\begin{equation}
 \Delta V_{1 loop}^{stop}= \frac{3}{32\pi^2}\left[\tilde{m}_{t_1}^4
 (\log \frac{\tilde{m}_{t_1}^2}{Q^2}- \frac{3}{2})+\tilde{m}_{t_2}^4
 (\log \frac{\tilde{m}_{t_2}^2}{Q^2}- \frac{3}{2})-2m_t^4
 (\log \frac{m_t^2}{Q^2}- \frac{3}{2})\right].
\end{equation}
Contributions from the other particles are much smaller~\cite{radcorr,G,carena}.

These corrections lead to the following modification of  the tree
level relation for the lightest Higgs mass
\begin{equation}
m_h^2 \approx M_Z^2\cos^2 2\beta + \frac{3g^2 m_t^4}{16\pi^2M_W^2}
\log\frac{ \tilde{m}_{t_1}^2\tilde{m}_{t_2}^2}{m_t^4}. \label{rad}
\end{equation}
One finds that the one loop correction is positive and increases the mass value.
Two loop  corrections have the opposite effect but are smaller and result
in slightly lower value of  the Higgs mass.

To find out  numerical values of these
corrections one has to determine the masses of all superpartners.
This means that one has to know the initial conditions for the
soft papameters. Fortunately due to the IRQFP solutions the
dependence on initial conditions may disappear at low energies.
This allows one to reduce the number of unknown parameters and
make predictions.

Due to extreme importance of the Higgs mass predictions for
experimental searches this problem has been the subject of intense
investigation. There is a considerable amount of papers devoted to
this  topic (see e.g. \cite{G,carena,Higgsmass,14,pl,hollik}).  And though the initial
assumptions and the strategy are different, the general
conclusions are very similar.

In what follows we accept the  strategy advocated in Refs.~\cite{YJK,JK}: As input
parameters one takes the known values of the top-quark, bottom-quark and
$\tau$-lepton masses, the experimental values of the gauge couplings, and
the mass of the Z-boson \cite{SM}. To reduce the arbitrariness of the soft terms we
use the fixed-point values for the Yukawa couplings and SUSY breaking
parameters.
 The value of $\tan\beta$  is determined from the relations
between the running quark masses and the Higgs v.e.v.'s in the MSSM
\begin{eqnarray} m_t &=& y_t \ v \ \sin\beta\,,
\label{mt} \\
m_b &=& y_b\  v \ \cos\beta\,, \label{mb} \\
m_{\tau} &=& y_{\tau}\  v \ \cos\beta\,. \label{mtau} 
\end{eqnarray}
The Higgs mixing parameter $\mu$ is defined from the minimization conditions
for the Higgs potential and requirement of radiative EWSB.

As an output we determine the mass spectrum of superpartners and of the
Higgs bosons as functions of the only free parameter, namely $m_{1/2}$,
which is directly related to the gluino mass $M_3$. Varying this parameter
within the experimentally allowed range, one gets all the masses as
functions of this parameter (see Table\ref{out}).

\begin{table}[ht]
\begin{center}
\begin{tabular}{|l|c|l|}
\hline
 \ \ \ \   INPUT &  & \ \ \ \  OUTPUT \\  \hline
$m_t(pole) =173.8\pm 5.2$ GeV &  &  $Y_t, Y_b, Y_\tau$\\
$m_b(pole)=4.94 \pm 0.15$ GeV & & $A_t, A_b, A_\tau$ \\
$m_\tau(pole) =1.7771\pm 0.0005$ GeV && $\tan\beta,  \mu$ \\
 $\alpha_3(M_Z)=0.118\pm 0.005$ &
$\Rightarrow$ & $\tilde{m}^2_Q, \tilde{m}^2_U,\tilde{m}^2_D$
 \\

$\alpha_2(M_Z)=0.034$ & & $\tilde{m}^2_L, \tilde{m}^2_E$
\\ $\alpha_1(M_Z)=0.017$ & & $m^2_{H_1}, m^2_{H_2}$ \\ $M_Z =91.187\pm 0.007$ GeV &
& $m_h, m_A, m_H, m_H^{\pm}$\\ \hline
\end{tabular} \end{center}
\caption{The values of input parameters and the output of fixed
point analysis }\label{out}
\end{table}

For low $tan\beta$ the value of $\sin\beta$ is determined from
eq.(\ref{mt}), while for high $\tan\beta$ it is more convenient to
use the relation $\tan\beta =\frac{m_t}{m_b}\frac{y_b}{y_t}$,
since the ratio $y_t/y_b$ is almost a constant in the range of
possible values of $y_t$ and $y_b$.

For the evaluation of  $\tan\beta$ one first needs to determine
the running top- and bottom-quark masses. One can find them using
the well-known relations to the pole masses (see e.g.
\cite{pole,P,14}), including both QCD and SUSY corrections. For
the top-quark one has:
\begin{equation}
m_t(m_t)=\frac{m_t^{pole}}{1+ \left(\frac{\Delta
m_t}{m_t}\right)_{QCD} + \left(\frac{\Delta
m_t}{m_t}\right)_{SUSY}}\ . \label{mtpole}
\end{equation}
 Then,  the
following procedure is used to evaluate the running top mass.
First, only the QCD correction is taken into account and
$m_t(m_t)$ is found in the first approximation. This allows one to
determine both the stop masses and the stop mixing angle. Next,
having at hand the stop and gluino masses, one takes into account
the stop/gluino corrections.

For the bottom quark the situation is more complicated because the
mass of the bottom quark $m_b$ is essentially smaller than the
scale $M_Z$ and so one has to take into account the running of
this mass from the scale $m_b$ to the scale $M_Z$. The procedure
is the following \cite{P,26,27}: one starts with the bottom-quark
pole mass, $m_b^{pole}=4.94 \pm 0.15$ \cite{28} and finds the SM
bottom-quark mass at the scale $m_b$ using the two-loop $QCD$
corrections
\begin{equation}
m_b(m_b)^{SM}=\frac{m_b^{pole}}{1+ \left( \frac{\Delta m_b}{m_b}
\right)_{QCD}}\,.  \label{mbpole}
\end{equation}
 Then, evolving this mass to the scale $M_Z$  and using a numerical
solution of the two-loop  SM RGEs \cite{P,27} with
$\alpha_3(M_Z)=0.12$ one obtains $m_b(M_Z)_{SM}=2.91$ GeV. Using
this value one can  calculate the sbottom masses and then return
back to take into account the SUSY corrections from massive SUSY
particles
\begin{equation}
m_b(M_Z)= \frac{m_b(M_Z)^{SM}}{1+ \left( \frac{ \Delta m_b }{ m_b
} \right)_{SUSY} }\,.
 \label{mbsusy}
\end{equation}

When calculating the stop and sbottom masses one needs to know the
Higgs mixing parameter $\mu$.  For determination of this parameter
one uses the relation between the $Z$-boson mass and the
low-energy values of $m_{H_1}^2$ and $m_{H_2}^2$ which comes from
the minimization of the Higgs potential:
\begin{equation}
\frac{M_Z^2}{2}+\mu^2=\frac{m_{H_1}^2+\Sigma_1-
(m_{H_2}^2+\Sigma_2) \tan^2\beta}{\tan^2\beta-1}\,, \label{MZC}
\end{equation}
where $\Sigma_1$ and $\Sigma_2$ are the radiative
corrections~\cite{G}. Large contributions to these functions come
from stops and sbottoms. This equation allows one to obtain the
absolute value of $\mu$, the sign of $\mu$ remains a free
parameter.

Whence the quark running masses and the $\mu$ parameter are found,
one can determine   the corresponding values of $\tan\beta$ with
the help of eqs.(\ref{mt},\ref{mb}). This gives in low and high
$\tan\beta$ cases, respectively \cite{YJK,JK,OK}
\begin{eqnarray*}
\tan{\beta}&=&1.47 \pm 0.15 \pm 0.05 \ \ \ for \ \mu>0\,,\\
 \tan{\beta}&=&1.56 \pm 0.15 \pm 0.05 \ \ \ for \ \mu<0\,,\\
\tan{\beta}&=&69.3 \pm 0.6 \pm 0.3 \ \ \ \ \ for \ \mu>0\,,\\
 \tan{\beta}&=&38.1 \pm 0.9 \pm 0.4 \ \ \ \ \ for \ \mu<0\,.
\end{eqnarray*}
The deviations  from the central value are connected with the
experimental uncertainties of the top-quark mass, $\alpha_3(M_Z)$
and uncertainty due to the fixed point values of $y_t(M_Z)$ and
$y_b(M_Z)$.

Having all the relevant parameters at hand it is possible to estimate the
masses of the Higgs bosons. With the fixed point behaviour  the only
dependence left is on $m_{1/2}$ or the gluino mass $M_3$. It is restricted
only experimentally: $M_3>144$ GeV \cite{EWWG} for arbitrary values of the
squarks masses.

Let us start with low $\tan\beta$ case. The masses of CP-odd,
charged and CP-even heavy Higgses increase almost linearly with
$M_3$. The main restriction comes from the experimental limit on
the lightest Higgs boson mass. It excludes $\mu <0$ case and for
$\mu>0$ requires the heavy gluino mass $M_3 \geq 750$ GeV.
Subsequently one obtains \cite{YJK}
 $$m_A >844\ GeV, \ \ m_{H^\pm}>846 \ GeV, \ \ m_H
> 848\  GeV, \ \ \  for  \ \mu>0,$$
i.e. these particles are too heavy to be detected in the nearest
experiments.

For high $\tan\beta$ already the requirement of positivity of
$m_A^2$ excludes the region with small $M_3$. In the most
promising region  $M_3>1$ TeV ($m_{1/2}>300$ GeV) for the both
cases $\mu>0$ and $\mu<0$ the masses of CP-odd, charged and
CP-even heavy Higgses are also too heavy to be detected in the
near future \cite{JK}
 $$m_A>1100 \ \mbox{GeV  for} \ \
\mu>0, \ \ \ m_A>570 \ \mbox{GeV for}\ \ \mu<0,$$
 $$m_{H^{\pm}} > 1105 \
\mbox{GeV for} \ \  \mu>0, \ \ \   m_{H^{\pm}}> 575\ \mbox{GeV
for}\ \ \mu<0.$$
 $$m_H > 1100 \ \mbox{GeV for} \ \  \mu>0, \ \ \ m_H > 570 \
\mbox{GeV  for}
 \ \ \mu<0.$$

The situation is different for the lightest Higgs boson $h$, which
is much lighter.    Radiative corrections  in this case are
crucial and increase the value of the Higgs mass substantially
\cite{carena, G}. They have been calculated up to the second loop
order \cite{hemp, hollik}. As can be seen from eq.(\ref{rad}) the
one loop correction is positive increasing the tree level value
almost up to 100\% and the second one is negative thus decreasing
it a little bit. We use in our analysis the leading two-loop
contributions evaluated in ref.\cite{carena}.  As has been already
mentioned these corrections depend on the masses of  squarks and
the other superpartners for  which we substitute the values
 obtained  above from the IRQFP's.

The results depend on the sign of  parameter $\mu$.  It is not
fixed since the requirement of  EWSB defines only the value of
$\mu^2$.   However, for low $\tan\beta$ negative values of $\mu$
lead to a very small  Higgs  mass which is already excluded by
modern experimental data, so further on we consider only the
positive values of $\mu$.  Fot high $\tan\beta$ both signs of $\mu$
are allowed.

\begin{figure}[ht]
\vspace{-7cm}
  \begin{center}
\epsfig{figure=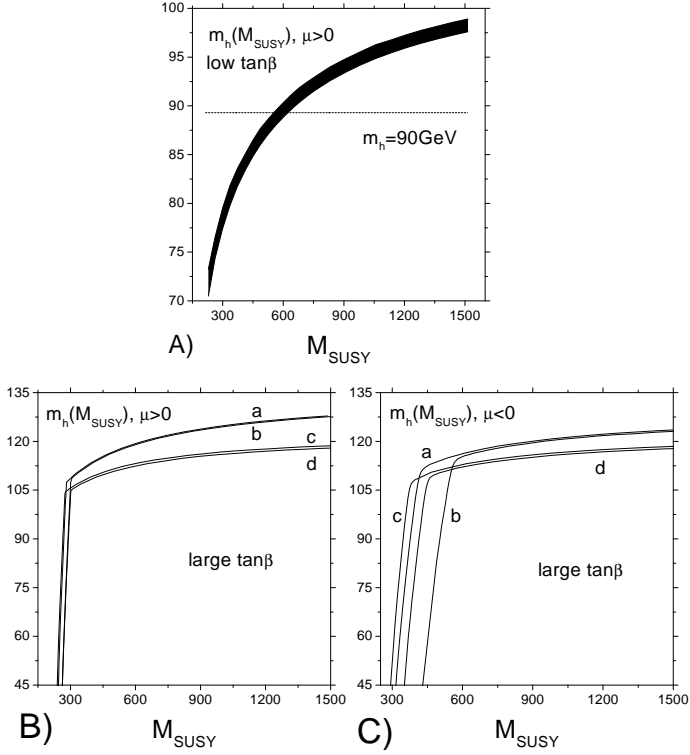, height=16cm}
  \end{center}
\caption{ A) The dependence of the mass of the lightest Higgs
boson $h$ on $M_{SUSY}=(\tilde m_{t_1} \tilde m_{t_2})^{1/2}$
(shaded area) for $\mu>0$, low $\tan\beta$.  The dashed line
corresponds to the minimum value of $m_h=90$ GeV allowed by
experiment.   B),C) The mass of the lightest Higgs boson $h$ as
function of $M_{SUSY}$ for different signs of $\mu$, large
$\tan\beta$. The curves (a,b) correspond to the upper limit of the
Yukawa couplings and to $m_0^2/m_{1/2}^2=0$ (a) or to
$m_0^2/m_{1/2}^2=2$ (b). The curves (c,d) correspond to the lower
limit of the Yukawa couplings and  to $m_0^2/m_{1/2}^2=0$ (c) or
 to $m_0^2/m_{1/2}^2=2$ (d).  Possible values of the mass of the
lightest Higgs boson are inside the areas marked by these lines.}
\label{fig:4}
\end{figure}

Consider first the low $\tan\beta$ regime.
At the upper part of Fig.\ref{fig:4} it is shown the value of
$m_h$ for $\mu>0$ as a function of the geometrical mean of stop
masses - this parameter is often identified  with  a supersymmetry
breaking scale $M_{SUSY}$. One can see that the value of $m_h$
quickly saturates close to $\sim$ 100 GeV. For $M_{SUSY}$ of the
order of 1 TeV the value of the lightest Higgs mass is~\cite{YJK}
\begin{equation}
m_h=(94.3+1.6+0.6\pm5\pm0.4) \ \mbox{GeV}, \ \ \ \mbox{ for} \
M_{SUSY}=1 \ TeV, \label{masshig}
\end{equation}
 where the first uncertainty comes from the deviations from the IRQFPs
 for the mass parameters, the second one is related to that of the
top-quark Yukawa coupling, the third reflects the uncertainty of
the top-quark mass of 5 GeV, and the last one comes from that of
the strong coupling.

One can see that the main source of uncertainty is the
experimental error in the top-quark mass. As for the uncertainties
connected with the fixed points, they give much smaller errors of
the order of 1 GeV.

The  obtained result (\ref{masshig}) is very close to the upper
boundary, $m_h=97$ GeV, obtained in Refs. \cite{14,pl}. Note,
however, that the uncertainties mentioned above as well as the
upper bound \cite{14,pl} are valid for the universal boundary
conditions. Loosing these requirement leads to increase of the
upper bound values of the Higgs mass in case of low $\tan\beta$ up
to $\sim 102$ GeV \cite{CJK}.

For the high $\tan\beta$ case the lightest Higgs is slightly
heavier, but the difference is crucial for LEP II. The mass of the
lightest Higgs boson as a function of   $M_{SUSY}$ is shown in
the lower part of
Fig.\ref{fig:4} . One has  the following values of $m_h$ at a
typical scale $M_{SUSY}=1$ TeV ($M_3 \approx 1.3$TeV)~\cite{JK}:
\begin{eqnarray}
m_h&=&128.2 -0.4 - 7.1 \pm 5 \ \mbox{GeV, \ for} \  \mu>0 \,,
\nonumber \\ m_h&=&120.6 -0.1 - 3.8 \pm 5 \ \mbox{GeV,\ for} \
\mu<0 \,.  \nonumber
\end{eqnarray}
The first uncertainty is connected with the deviations from the
IRQFPs for mass parameters, the second one with the Yukawa
coupling IRQFPs, and the third one is due to the experimental
uncertainty in the top-quark mass. One can immediately see that
the deviations from the IRQFPs for mass parameters are negligible
and only influence the steep fall of the function on the left,
which is related to the restriction on the CP-odd Higgs boson mass
$m_A$.  In contrast with the low $\tan\beta$ case, where the
dependence on the deviations from Yukawa fixed points was about
$1$ GeV, in the present case it is much stronger. The experimental
uncertainty in the strong coupling constant $\alpha_s$ is not
included  because it is negligible compared to those of the
top-quark mass and the Yukawa couplings and is not essential here
contrary to the low $\tan\beta$ case.

One can see that for large $\tan\beta$ the masses of the lightest
Higgs boson are typically around 120 GeV that is too heavy for
observation at LEP II. Note, however, that the uncertainties
increase if one considers the non-universal boundary conditions
which decreases the lower boundary for the Higss mass\footnote{Lep
II is now increasing its energy and may possibly reach the lower
bound of high $\tan\beta$ scenario predictions.}. At the same time
recent experimental data has practically excluded the low
$\tan\beta$ scenario and in the next year the situation will be
completely clarified.

Thus, one can see that in the IRQFP approach all the Higgs bosons
except for the lightest one are found to be too heavy to be
accessible in the nearest experiments. This conclusion essentially
coincides with the results of more sophisticated analyses. The
lightest neutral Higgs boson, on the contrary is always light. The
situation may improve a bit if one considers more complicated
models with enlarged Higgs structure \cite{ext}. However, it does
not change the generic feature of SUSY theories, the presence of
light Higgs boson.

In the near future with the operation of the Tevatron and LHC
proton-antiproton accelerators with the c.m energy of 2 and 15
TeV, respectively, the energy range from 100 to 1000 GeV will be
scanned and  supersymmetry will be either discovered or the
minimal version will be abandoned in particle physics. The Higgs
boson might be the first target in search of SUSY.

\section {Conclusion}

We have attempted to show how following the pattern of RG flow one
can explore physics lying beyond the Standard Model of fundamental
interactions. The methods are essentially based on renormalization
group technique in the framework of quantum field theory. The
running of parameters is the key ingredient in attempts to go
beyond the range of energies accessible to modern accelerators, to
find manifestation of new physics. In the absence of solid
experimental facts the RG flow  happens to be the only sauce of
definite predictions. The peculiarity of the moment is that these
predictions are the subject of experimental tests today, and in
the near future, so one may hope to check the  ideas described
above at the turn of the Millennium.

\vglue 0.5cm
 {\bf Acknowledgments}

\vglue 0.5cm

The author is grateful to S.Codoban, A.Gladyshev,
M.Jur\v{c}i\v{s}in, V.Velizhanin, K.Ter-Martirossian, R.Nevzorov,
W. de Boer and G.Moultaka for useful discussions and to S.Codoban
for help in preparing the manuscript.  Special thanks to the organizers
of the conference D.O'Connor and C.Stephens.  Financial support from RFBR
grants \# 99-02-16650 and \# 96-15-96030   is kindly acknowledged.

\end{document}